\documentclass[aps,prd,twocolumn,superscriptaddress,showkeys,showpacs]{revtex4}
\usepackage{pdfsync}
\usepackage{mathptmx}
\usepackage[utf8]{inputenc}
\usepackage[T1]{fontenc}
\usepackage{slashed}
\usepackage{times}
\usepackage{amssymb,amsfonts,amsmath,amsthm}
\usepackage{dsfont,bbm}
\usepackage{dcolumn}
\usepackage{epsf}
\usepackage{graphicx}
\usepackage[caption=false]{subfig}
\usepackage{dsfont}
\usepackage{slashed}
\usepackage[active]{srcltx}
\usepackage[usenames]{color}
\usepackage{simplewick}

\begin{document}

\title{Drell-Yan hadron tensor: contour gauge and gluon propagator}

\author{I.~V.~Anikin}
\email{anikin@theor.jinr.ru}
\affiliation{Bogoliubov Laboratory of Theoretical Physics, JINR,
             141980 Dubna, Russia}
\author{I.~O.~Cherednikov}
\email{igor.cherednikov@uantwerpen.be}
\affiliation{SCK-CEN, B-2400 Mol, Belgium}
\affiliation{Departement Fysica, Universiteit Antwerpen,
             B-2020 Antwerpen, Belgium}
\author{O.~V.~Teryaev}
\email{teryaev@theor.jinr.ru}
\affiliation{Bogoliubov Laboratory of Theoretical Physics, JINR,
             141980 Dubna, Russia}

\begin{abstract}
We consider the gauge invariant Drell-Yan hadron tensor which includes the
standard and non-standard diagram contributions.
The non-standard diagram contribution is appeared owing to
the complexity of the twist three $B^V(x_1,x_2)$-function where
the gluon pole manifests.
We use the contour gauge conception which allows us to fix easily
the spurious uncertainties in the gluon propagator.
The contour gauge condition is generated by the corresponding Wilson lines
in both the standard and non-standard diagrams. We demonstrate the substantial
role of the non-standard diagram for forming of the relevant contour in the Wilson
path-ordered exponential that leads to the spurious singularity fixing.
\end{abstract}
\date{\today}
\maketitle

\section{Introduction}

The investigation of nucleon (hadron) composite structure is still the most important subjects of hadron physics.
From the experimental point of view, one of the wide-spread and useful instruments for such studies is
the single spin asymmetry (SSA). Especially, the single transverse spin asymmetry
opens the access to the three-dimensional nucleon structure thanks for
the non-trivial connection between the transverse spin and the parton transverse momentum dependence
(see, for example, \cite{Angeles-Martinez:2015sea, Boer:2011fh, Boer:2003cm, Kang:2011hk, Boer:2011fx})

In QCD, the SSA related to the Drell-Yan (DY) process was first considered in the case of
the longitudinally polarized hadron \cite{P-R,Carlitz:1992fv}.
This SSA is especially
interesting provided the second hadron is a pion. This is because of the sensitivity
\cite{Brandenburg:1995pk, Bakulev:2007ej}
to the shape of pion distribution amplitude, being currently the object of major interest
\cite{Radyushkin:2009zg,Polyakov:2009je}
(see also \cite{Mikhailov:2009sa} and the references therein).
It was shown that the imaginary phase in the SSA which is associated with the longitudinally polarized nucleon
appear due to either the hard perturbative gluon loops \cite{P-R,Carlitz:1992fv}
or twist four contribution of the  pion distribution amplitude
\cite{Brandenburg:1995pk,Bakulev:2007ej}.

For the single transverse spin asymmetry in the transverse-polarized DY process,
the imaginary part has previously been extracted from the quark propagator in
the so-called standard, see Fig.~\ref{Fig-DY} the left panel,
diagram with quark-gluon twist three correlator only
(it leads  to the gluon pole contribution to SSA, see \cite{Hammon:1996pw, Boer:1997bw}).
The reason was that the ambiguity in the boundary conditions provide the purely real
quark-gluon function $B^V(x_1,x_2)$ which parameterizes
$\langle\bar\psi\gamma^+A^\perp\psi\rangle$ matrix element.
On the other hand, the real $B^V(x_1,x_2)$-function kills the contribution from the
non-standard, see Fig.~\ref{Fig-DY} the right panel, diagram which however is absolutely necessary
to ensure the QED gauge invariance of the DY hadron tensor. This situation has been discussed in detail
in series of papers \cite{AT-GP} where, with the help of the contour gauge conception,
the twist three $B^V(x_1,x_2)$-function has been proven to be in fact the complex function.
In its turn, this leads to the non-zero contribution from the non-standard diagram which
produces the imaginary phase required to have the SSA. This additional contribution
also leads to an extra factor of $2$ for SSA.

Recently, the problem of the spurious singularity fixing in the (local) axial gauge
has attracted an attention again (see, for example, \cite{Boer:2003cm, Belitsky:2002sm, Chirilli:2015fza}).

The light-cone axial gauge condition imposed on non-Abelian gluon field, $A^+=0$,
naturally enables the parton number (probability) interpretation of parton density functions in the tree level
\cite{Soper:1979fq, Collins:1981uw}. However, perturbative calculations beyond the tree
approximation demand careful treatment of the so-called spurious uncertainties in the gluon propagator
$D_{\mu\nu} (k)$ in the light-cone gauge \cite{Leibbrandt:1987qv, Slavnov:1987yh, Bassetto:1991ue, Bassetto:1984dq}.
The latter arise as ill-defined pole singularities of the form  $\sim (k^+)^{-1}$ and are associated, putting
the same issue a bit different, with the residual gauge freedom due to incomplete gauge fixing
by $A^+=0$. For this reason calculations in the axial (light-cone) gauge in higher perturbative orders
are cumbersome and sometimes even contradictory \cite{Collins:1989gx, Ivanov:1990vy}.
One can attempt to overcome this difficulty by working in the well-defined general
covariant gauge setting the gauge parameter to
$\xi = - 3 + 0 (\alpha_s)$,
which is known to effectively `imitate' non-covariant gauges \cite{Stefanis:1983ke, Ivanov:1990vy}.
Another approach is to keep working in the light-cone gauge and to get rid of the residual gauge freedom
by an appropriate extra gauge-fixing condition.
The latter can be obtained in terms of the various boundary conditions for the gluon fields and/or
their spatial derivatives \cite{Belitsky:2002sm, Boer:2003cm, Chirilli:2015fza}.

In the present work, we investigate an alternative approach to formulation of the more general
gauge-fixing condition from the very beginning which is supposed to entail the ``right'' pole
prescriptions for the gluon propagator.
We continue to explore the contour gauge conception and
demonstrate how the spurious uncertainties in the gluon propagator can ultimately be fixed
in the non-local axial gauges.
Working within the collinear factorization procedure, we
emphasize the substantial role of the non-standard diagram
to get the relevant contour in the Wilson
path-ordered exponential needed to fix ultimately the spurious singularity in the gluon propagator.

\section{Getting started: kinematics}

We begin with the kinematics of Drell-Yan process.
As in \cite{AT-GP}, we study the Drell-Yan process with the transversely polarized hadron:
\begin{eqnarray}
N^{(\uparrow\downarrow)}(p_1) + N(p_2) &\to& \gamma^*(q) + X(P_X)
\nonumber\\
&\to&\ell(l_1) + \bar\ell(l_2) + X(P_X),
\end{eqnarray}
where the virtual photon producing the lepton pair ($l_1+l_2=q$) has a large mass squared
($q^2=Q^2$) while the transverse momenta are small and integrated out.
This kinematics (anticipating the collinear factorization procedure) suggests a convenient frame with
fixed dominant light-cone directions \cite{AT-GP}:
\begin{eqnarray}
\label{kin-DY}
&&p_1\approx \frac{Q}{x_B \sqrt{2}}\, n^*\, , \quad p_2\approx \frac{Q}{y_B \sqrt{2}}\, n,
\\
&&n^{*\,\mu}=\big(\frac{1}{\sqrt{2}},\,\vec{\bf 0}_\perp,\,\frac{1}{\sqrt{2}}\big)=
\big(1^+,0^-,\vec{\bf 0}_\perp \big),
\nonumber\\
&&n^{\mu}=\big(\frac{1}{\sqrt{2}},\,\vec{\bf 0}_\perp,\,\frac{-1}{\sqrt{2}}\big)=
\big(0^+,1^-,\vec{\bf 0}_\perp \big),
\nonumber\\
&&n^*\cdot n=n^{*\,+}n^{-}=1
\nonumber
\end{eqnarray}
It is also instructive to introduce the dimensionful analogs of $n, n^*$ as
\begin{eqnarray}
\label{ntilde}
\tilde n^-=\frac{p_2^-}{p_1p_2},\quad \breve n^+=\frac{p_1^+}{p_1p_2}.
\end{eqnarray}
With the above vectors as a basis, an arbitrary vector can be (Sudakov) decomposed as
\begin{eqnarray}
\label{Sud-Exp}
&&a^{\mu} = a^+ n^{*\,\mu} + a^- n^{\mu} + a^\mu_\perp,
\nonumber\\
&&a^{\mu,+}\stackrel{{\rm def}}{=} a^+ n^{*\,\mu},
\quad a^{\mu,-}\stackrel{{\rm def}}{=} a^- n^{\mu}.
\end{eqnarray}
In what follows we will not be so precise about writing the covariant and contravariant vectors in any kinds of
summations over the four-dimensional vectors, except the cases where this trick may lead to misunderstanding.

\section{Drell-Yan hadron tensor: Derivation of Wilson lines}

The polarized DY process is very convenient process to study the role of twist three by exploring
of different kinds of SSAs.
For example, one can study the left-right asymmetry which means the transverse momenta
of the leptons are correlated with the direction
$\textbf{S}\times \textbf{e}_z$ where $S_\mu$ implies the
transverse polarization vector of the nucleon and $\textbf{e}_z$ is a beam direction \cite{Barone}.

Generally speaking, any single spin asymmetries can be presented in the symbolical form as
\begin{eqnarray}
\label{SSA-A}
{\cal A} \sim d\sigma^{(\uparrow)} - d\sigma^{(\downarrow)}
\sim {\cal L}_{\mu\nu}\, {\cal W}_{\mu\nu}\, ,
\end{eqnarray}
where ${\cal L}_{\mu\nu}$ is an unpolarized leptonic tensor and
${\cal W}_{\mu\nu}$ stands for the hadronic tensor. At the moment,
we do not specify the phase space in Eqn.~(\ref{SSA-A}) because the exact expression for SSA is irrelevant
for our discussion. Instead, we mainly pay our attention on the hadron tensor which can be presented as
\begin{eqnarray}
\label{W-over-g}
{\cal W}_{\mu\nu}&&\hspace{-.3cm}={\cal W}^{(0)}_{\mu\nu} + {\cal W}^{(1)}_{\mu\nu}(g|A) +
{\cal W}^{(2)}_{\mu\nu}(g|A) + (g^n\text{-terms}|n\geq 2)
\nonumber\\
&&\hspace{-.3cm}=
\overline{{\cal W}}^{(0)}_{\mu\nu}(A^{\pm}) +{\cal W}^{(1)}_{\mu\nu}(g|A^\perp) +
{\cal W}^{(2)}_{\mu\nu}(g|A^\perp)+ \cdots,
\end{eqnarray}
where $g$ denotes the strong interaction coupling constant and
\begin{eqnarray}
\label{barW}
\overline{{\cal W}}^{(0)}_{\mu\nu}(A^{\pm})=
{\cal W}^{(0)}_{\mu\nu} + {\cal W}^{(1)}_{\mu\nu}(g|A^+) +
{\cal W}^{(2)}_{\mu\nu}(g|A^-) + \cdots\,.
\end{eqnarray}
The hadron tensor representations can be found below.
In our case, the single transverse spin asymmetry is only generated by the hadron tensors
${\cal W}^{(1)}_{\mu\nu}(g|A^\perp)$ and ${\cal W}^{(2)}_{\mu\nu}(g|A^\perp)$ where
the twist three contributions related to $\langle\bar\psi\gamma^+A^\perp\psi\rangle$
have been extracted.
As shown below, the $\langle\bar\psi\gamma^+A^\pm\psi\rangle$-correlators
in the hadron tensors ${\cal W}^{(1,2)}_{\mu\nu}(g|A)$
participate in forming of the corresponding Wilson lines which appear in the
quark-antiquark correlators of the hadron tensor $\overline{{\cal W}}^{(0)}_{\mu\nu}(A^{\pm})$.
In the frame of usual axial gauge ($A^+=0$), this kind of contributions can be
discarded.
However, we work in the contour gauge which is, first, a non-local generalization of the well-know axial gauge.
Second, the contour gauge contains the important and unique additional information (needed to fix
the prescription in the gluon poles) which is invisible in the case of usual (local) axial gauge.
From this point of view, before we discard the terms with $A^+$, we have to determine
the relevant fixed path in the restored Wilson line with $A^+$
which eventually leads to the certain prescriptions in the gluon poles (for further explanations, see \cite{AT-GP}).

\subsection*{The standard hadron tensor (direct process)}

\begin{widetext}
\begin{figure*}[ht]
\centerline{\includegraphics[width=0.45\textwidth]{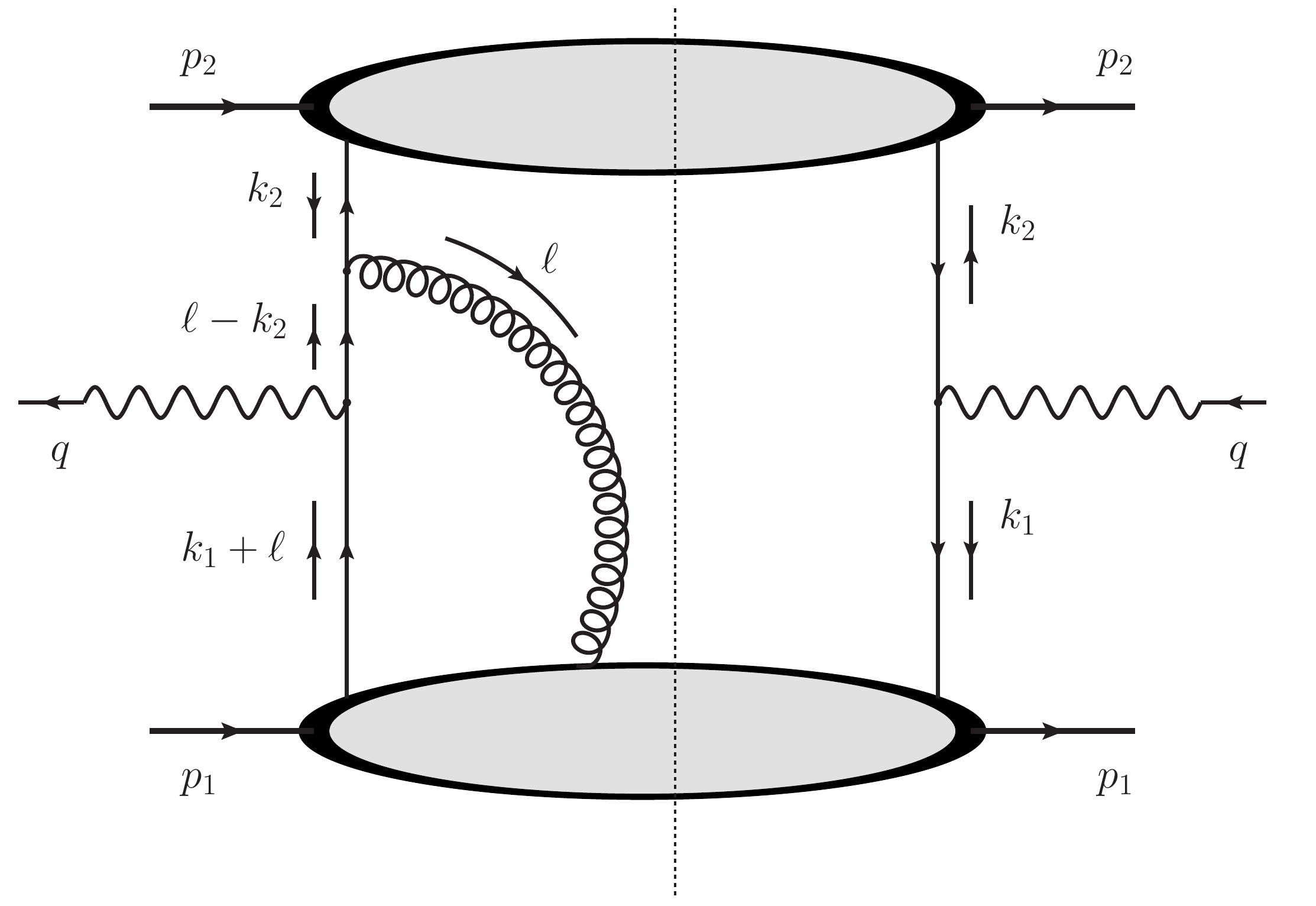}
\hspace{0.2cm}
\includegraphics[width=0.45\textwidth]{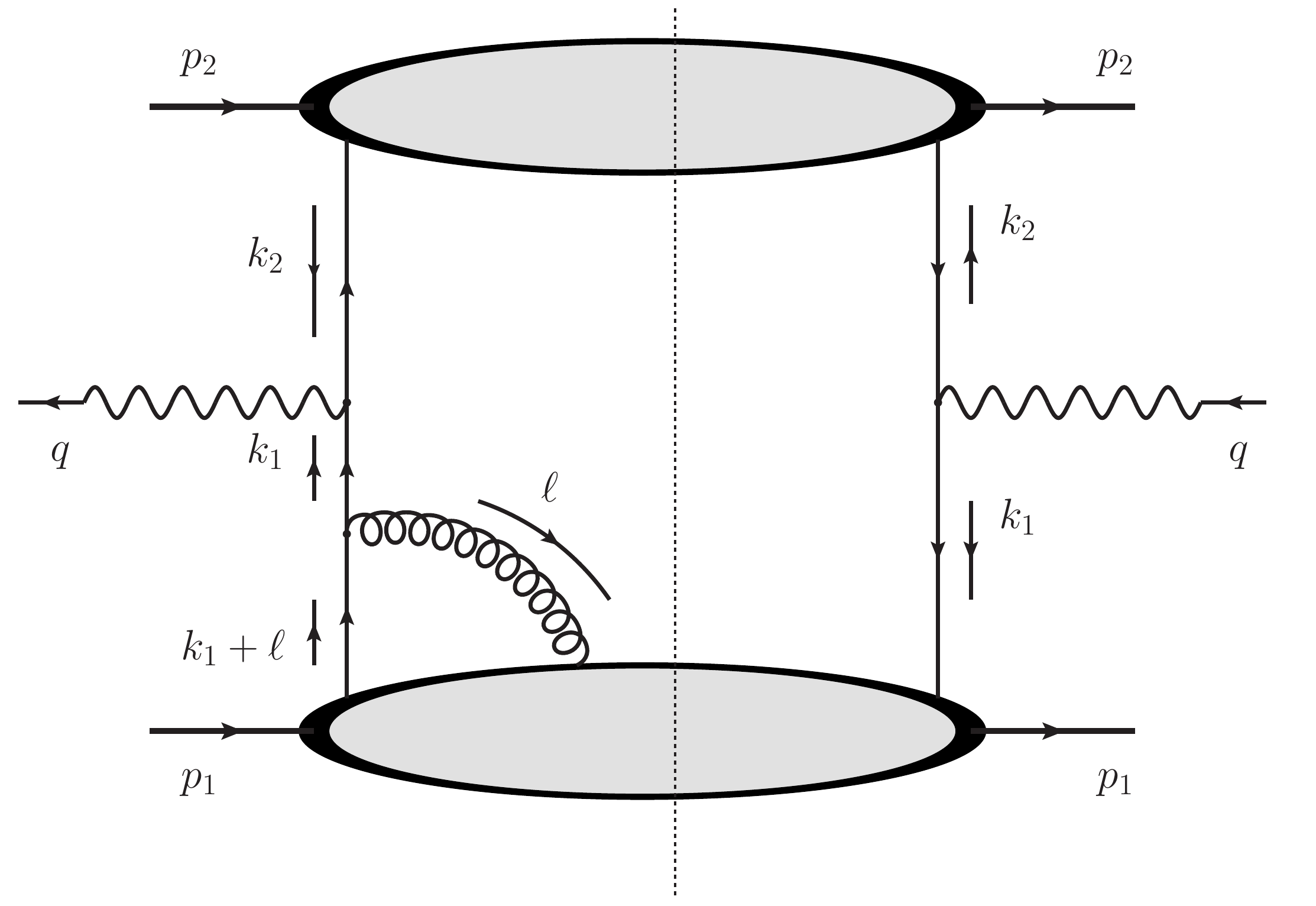}}
  \caption{The Feynman diagrams which contribute to the polarized Drell-Yan hadron tensor:
  the standard (the left panel) and non-standard diagrams (the right panel).}
\label{Fig-DY}
\end{figure*}
\end{widetext}

In this section, we analyse the part of the DY hadron tensor which is generated by the diagram
in Fig.~\ref{Fig-DY}, the left panel. This is the standard hadron tensor which can be written
in non-factorized form as
\begin{eqnarray}
\label{HadTen1-2}
&&{\cal W}^{(1)}_{\mu\nu}(g|A)=\int d^4 k_1\, d^4 k_2 \, \delta^{(4)}(k_1+k_2-q)
\, \bar\Phi^{[\gamma^-]} (k_2) \times
\nonumber\\
&&\int d^4 \ell \,
\Phi^{(A)\,[\gamma^+]}_\alpha (k_1,\ell) \,
\text{tr}\Big[
\gamma_\mu  \gamma^- \gamma_\nu \gamma^+ \gamma_\alpha\times
\nonumber\\
&&
\frac{(\ell^+-k_2^+)\gamma^- + (\ell^- - k_2^-)\gamma^+ -
(\vec{\ell}_\perp - \vec{k}_{2\,\perp})\vec{\gamma}_\perp}
{(\ell - k_2)^2 + i\epsilon}
\Big] \, ,
\end{eqnarray}
where
\begin{eqnarray}
\label{PhiF-1}
&&\hspace{-0.8cm}\Phi^{(A)\,[\gamma^+]}_\alpha (k_1,\ell)
\stackrel{{\cal F}_2}{=}
\langle p_1, S^T | \bar\psi(\eta_1)\gamma^+  gA_{\alpha}(z)  \psi(0) | S^T, p_1\rangle ,
\\
\label{PhiF-2}
&&\hspace{-0.8cm}\bar\Phi^{[\gamma^-]}(k_2)\stackrel{{\cal F}_1}{=}
\langle p_2 | \bar\psi(\eta_2)\gamma^- \psi(0)| p_2\rangle .
\end{eqnarray}
In Eqns.~(\ref{PhiF-1}) and (\ref{PhiF-2}), ${\cal F}_1$ and  ${\cal F}_2$
denote the Fourier transformation with the measures defined as
\begin{eqnarray}
d^4\eta_2\, e^{ik_2\cdot\eta_2}\,\,\, \text{and} \,\,\,
d^4\eta_1\, d^4 z\, e^{-ik_1\cdot\eta_1-i\ell\cdot z} ,
\end{eqnarray}
respectively. For the sake of shortness, we will omit $S^T$ in
the hadron states that indicates the transverse polarization of hadron.
We now analyze the tensor structure of the trace in Eqn.~(\ref{HadTen1-2}).
We can see that the first term of the quark propagator, $\ell^+-k_2^+)\gamma^-$, singles out
only the transverse components of gluon field in the quark-gluon correlator, see Eqn.~(\ref{PhiF-1}).
At the same time, the second term of the quark propagator, $(\ell^- - k_2^-)\gamma^+$, separates out
only the longitudinal component $A^+$ in the quark-gluon correlator. This second term is very important for
derivation of the corresponding Wilson line which defines in our approach the contour gauge.
And, the third term of the quark propagator give us the quark-gluon correlator with both indices $\alpha=(+, \perp)$.

The collinear factorization procedure for the process under consideration can be introduced
by the following steps (for the details see, e.g.,
Refs. \cite{Efremov:1984ip, Anikin:2009bf}):

\noindent (a) the decomposition of loop integration momenta around the corresponding dominant direction:
$$k_i = x_i p + (k_i\cdot p)n + k_T$$
within the certain light cone basis formed by the vectors $p$ and $n$ (in our case, $n^*$ and $n$);

\noindent (b) the replacement:
$$d^4 k_i \Longrightarrow d^4 k_i \,dx_i \delta(x_i-k_i\cdot n)$$
that introduces the fractions with the appropriated spectral properties;

\noindent (c) the decomposition of the corresponding propagator products, which will finally form the hard part,
around the dominant direction. It is necessary to notice that
in the DY process case the corresponding $\delta$-functions appeared in the hadron tensor
and expressed the momentum conservation law should be also referred to the hard parts.
This statement was argued in \cite{An} in the context of the so-called factorization links;

\noindent (d) the use of the collinear Ward identity if it is necessary within the given of accuracy level;

\noindent (e) performing of the Fierz decomposition for $\psi_\alpha (z) \, \bar\psi_\beta(0)$ in
the corresponding space up to the needed projections.

Let us first dwell on the second term, $(\ell^- - k_2^-)\gamma^+$, contribution.
This term is responsible for forming of the Wilson line in the gauge-invariant quark-antiquark string operator.
Indeed, making used the collinear factorization ($\ell^-\approx 0, (\ell-k_2)^2\approx -2\ell^+k_2^-$),
the above-mentioned term contributes in the hadron tensor as
\begin{eqnarray}
\label{HadTen1-3}
&&{\cal W}^{(1)\, [k_2^-]}_{\mu\nu}(g|A^+)=\int d\mu (k_i;x_1,y) \,
\, \bar\Phi^{[\gamma^-]} (k_2)\frac{1}{2} \int dz^-
\times
\nonumber\\
&&
\text{tr}\Big[
\gamma_\mu  \gamma^- \gamma_\nu \gamma^+ \gamma^- \gamma^+
\Big]
 \int d\ell^+\, \frac{e^{ - i\ell^+ z^-}}{\ell^+ - i\epsilon}
\int d^4\eta_1  \, e^{-ik_1\cdot\eta_1}\times
\nonumber\\
&&
\langle p_1 | \bar\psi(\eta_1)\, \gamma^+ \, gA^+(0^+,z^-,\vec{{\bf 0}}_\perp) \, \psi(0)
|p_1\rangle \, ,
\end{eqnarray}
where the integration measure reads
\begin{eqnarray}
&&\hspace{-0.3cm}d\mu (k_i;x_1,y)= dx_1 d^4k_1 \delta\Big(x_1-\frac{k_1^+}{p_1^+}\Big)
\,dy d^4k_2 \delta\Big(y-\frac{k_2^-}{p_2^-}\Big)\times
\nonumber\\
&&\hspace{-0.3cm}\big[ \delta^{(4)}(x_1p_1+yp_2 - q) \big].
\end{eqnarray}
The prescription $-i\epsilon$ in the denominator of (\ref{HadTen1-3})
directly follows from the standard causal
prescription for the massless quark propagator in (\ref{HadTen1-2}) (cf. \cite{Braun}).

Integration over $\ell^+$ in (\ref{HadTen1-3}), using the well-known integral representation
\begin{eqnarray}
\label{theta-func}
\theta(\pm x) =\frac{\pm i}{2\pi} \int\limits_{-\infty}^{+\infty} dk\, \frac{e^{-ikx}}{k\pm i\epsilon},
\end{eqnarray}
leads to the following expression:
\begin{eqnarray}
\label{HadTen1-4}
&&\hspace{-0.4cm}{\cal W}^{(1)\,  [k_2^-]}_{\mu\nu}(g|A^+)= \int d\mu (k_i;x_1,y) \times
\\
&&\hspace{-0.4cm}\text{tr}\Big[
\gamma_\mu  \gamma^- \gamma_\nu \gamma^+
\Big] \, \bar\Phi^{[\gamma^-]} (k_2)\int d^4\eta_1 \, e^{-ik_1\cdot\eta_1} \times
\nonumber\\
&&\hspace{-0.4cm}
\langle p_1| \bar\psi(\eta_1)\, \gamma^+
ig \int\limits_{-\infty^-}^{0^-} dz^- A^+(0^+,z^-,\vec{{\bf 0}}_\perp)
 \psi(0) | p_1\rangle ,
\nonumber
\end{eqnarray}
where we use
\begin{eqnarray}
\frac{1}{2} \gamma^+ \gamma^- \gamma^+ = \gamma^+.
\end{eqnarray}
It is important to stress that the leading order hadron tensor ${\cal W}^{(0)}_{\mu\nu}(g^0)$ differs from the
hadron tensor (\ref{HadTen1-2}) by overall sign: the leading hadron tensor has a pre-factor $i^2$ due to
two photon vertices, while the hadron tensor (\ref{HadTen1-2}) is accompanying by a pre-factor $i^4$
thanks for two photon and one gluon vertices together with the pre-factor from
the massless quark propagator $(-1)/i$ (we use the convention as in \cite{BogoShir}).

Thus, if we include all gluon emissions from the antiquark going from the upper blob
in Fig.~\ref{Fig-DY}, the left panel, (the so-called initial state interactions),
we are able to get the corresponding $P$-exponential in
$\Phi^{(A)\,[\gamma^+]}_\alpha (k_1,\ell)$. The latter is now represented by the
following matrix element:
\begin{eqnarray}
\label{me-Pexp}
\int d^4\eta_1 \, e^{-ik_1\cdot\eta_1}
\langle p_1 | \bar\psi(\eta_1)\, \gamma^+ \, [-\infty^- ;\, 0^-]_{A^+}
\, \psi(0) | p_1\rangle \, ,
\end{eqnarray}
where
\begin{eqnarray}
\label{Pexp-1}
&&[-\infty^- ;\, 0^-]_{A^+}\equiv [0^+, -\infty^-, \vec{{\bf 0}}_\perp ;\, 0^+, 0^-, \vec{{\bf 0}}_\perp]_{A^+}=
\nonumber\\
&&\mathbb{P}{\rm exp}\Big\{ i g \int\limits^{-\infty^-}_{0^-}
 dz^- \, A^+(0^+,z^-,\vec{{\bf 0}}_\perp) \Big\}\, .
\end{eqnarray}
The collinear twist ($t=d-s_a$) of $A^+$ is equal to zero, therefore
the Wilson line which is summing up all these components does not affect the
twist expansion within the collinear factorization.

If now we include in our consideration the gluon emission from
the incoming antiquark (the mirror contributions), we will obtain the Wilson
line $[\eta_1^-,-\infty^-]$ which
will ultimately give us, together with (\ref{Pexp-1}), the Wilson line connecting
the points $0$ and $\eta_1$ in (\ref{me-Pexp}) contributing to
$\overline{{\cal W}}^{(0)}_{\mu\nu}$.
This is exactly what happens, say, in the spin-averaged DY process \cite{Efremov:1978xm}.
However, for the SSA, these
two diagrams should be considered individually.
Indeed, their contributions to SSAs, contrary to spin-averaged case,
differ in sign and the dependence on the boundary point at $-\infty^-$
does not cancel.

For the pedagogical reason, we want to show the exponentiation of the transverse gluon field
(here, we mainly follow to \cite{Belitsky:2002sm}),
although we are restricted by the twist three case and the inclusion of all degrees of the transverse gluon field
exceeds our accuracy.
Let us consider the third term, $(\vec{\ell}_\perp - \vec{k}_{2\,\perp})\vec{\gamma}_\perp$, contribution
which helps us to demonstrate the exponentiation of the transverse gluon fields.
The corresponding hadron tensor part takes the following form
\begin{eqnarray}
\label{HadTen1-4-1}
&&{\cal W}^{(1)\, [\vec{\ell}_\perp]}_{\mu\nu}(g|A^\perp)=
\\
&&\int d\mu (k_i;x_1,y) \,
\, \bar\Phi^{[\gamma^-]} (k_2)
\text{tr}\Big[
\gamma_\mu  \gamma^- \gamma_\nu \gamma^+ \gamma^\perp_\alpha \vec{\gamma}_i^\perp
\Big]\times
\nonumber\\
&& \int d^4\ell \, \frac{(\vec{\ell}^\perp - \vec{k}_2^\perp)_i}{2\ell^+k_2^- + \vec{\ell}_\perp^{\,2} - i\epsilon}
\Phi_\alpha^{(A^\perp)[\gamma^+]}(k_1,\ell)
\equiv
\nonumber\\
&& \int d\mu (k_i;x_1,y) \,
\, \bar\Phi^{[\gamma^-]} (k_2)
\text{tr}\Big[
\gamma_\mu  \gamma^- \gamma_\nu \gamma^+ \gamma^\perp_\alpha \vec{\gamma}_i^\perp
\Big] \mathfrak{L}_{i,\alpha}
\,,
\nonumber
\end{eqnarray}
where we assume that $\vec{k}_{2\,\perp}\approx 0$. In Eqn.~(\ref{HadTen1-4}) let us focus
on the $\ell$-integration, we have
\begin{eqnarray}
\label{HadTen-ell}
&&\mathfrak{L}_{i,\alpha}= \int d\ell^+d\ell^- d^2\vec{\ell}_\perp \,
\frac{\vec{\ell}^\perp_i}{2\ell^+k_2^- + \vec{\ell}_\perp^{\,2} - i\epsilon}\times
\\
&&\int d^4\eta_1\,d^4z\, e^{-ik_1\eta_1 - i \ell z}
\langle p_1| \bar\psi(\eta_1)\gamma^+  gA^\perp_{\alpha}(z) \psi(0) | p_1\rangle.
\nonumber
\end{eqnarray}
We now use the $\alpha$-representation for the denominator that stems from the
quark propagator:
\begin{eqnarray}
\label{alpha-reps}
\frac{1}{2\ell^+k_2^- + \vec{\ell}_\perp^{\,2} - i\epsilon}=
i\int\limits_0^\infty d\alpha\, e^{-i\alpha[2\ell^+k_2^- + \vec{\ell}_\perp^{\,2} - i\epsilon]}.
\end{eqnarray}
Next, in Eqn.~(\ref{HadTen-ell}) we perform the integrations over $d\ell^-$ and $d\ell^+$
which give $\delta(z^+)$ and $\delta(z^- + 2\alpha k_2^-)$, respectively.
We remind that the variables $\alpha$ in (\ref{alpha-reps}) are dimensionful and
$\text{dim}_M[\alpha]=-2$.

Therefore, the integral  $\mathfrak{L}$ takes the following form (cf. \cite{Belitsky:2002sm})
\begin{eqnarray}
\label{HadTen-ell-2}
&&\mathfrak{L}_{i,\alpha}= i \int d^2\vec{\ell}_\perp \,\vec{\ell}^\perp_i
\int\limits_0^\infty d\alpha\, e^{-i\alpha[\vec{\ell}_\perp^{\,2} - i\epsilon]}
\int d^4\eta_1 \, d^2 \vec{z}_\perp
\times
\\
&&e^{-ik_1\eta_1 + i \vec{\ell}_\perp \vec{z}_\perp}
\langle p_1| \bar\psi(\eta_1)\gamma^+  gA^\perp_{\alpha}(0^+, -\infty^-, \vec{z}_\perp) \psi(0) | p_1\rangle.
\nonumber
\end{eqnarray}
In Eqn.~(\ref{HadTen-ell-2}) the transverse gluon field operator can be presented as
\begin{eqnarray}
\label{At}
A^\perp_{\alpha}(0^+, -\infty^-, \vec{z}_\perp) = \frac{\partial}{\partial z^\perp_\alpha}
\int\limits_{\mathds{C}}^{z^\perp} d\omega^\perp_\beta A^\perp_\beta(0^+, -\infty^-, \vec{\omega}^\perp),
\end{eqnarray}
where we fix the arbitrary constant $\mathds{C}$ to be $-\vec{\infty}_\perp$.
By making use of the representation (\ref{At}), after integration over $\alpha$ we arrive at
\begin{eqnarray}
\label{HadTen-ell-3}
&&\hspace{-0.8cm}
\mathfrak{L}_{i,\alpha}= i \int d^2\vec{\ell}_\perp \,
\frac{\vec{\ell}^\perp_i \ell^\perp_\alpha}{\vec{\ell}_\perp^{\,2} - i\epsilon}
\int d^4\eta_1 \, d^2 \vec{z}_\perp
e^{-ik_1\eta_1 + i \vec{\ell}_\perp \vec{z}_\perp}\times
\nonumber\\
&&\hspace{-0.8cm}
\langle p_1| \bar\psi(\eta_1)\gamma^+
g \int\limits_{-\infty^\perp}^{z^\perp} d\omega^\perp_\beta A^\perp_\beta(0^+, -\infty^-, \vec{\omega}^\perp)
\psi(0) | p_1\rangle.
\end{eqnarray}
We insert the obtained expression for $\mathfrak{L}_{i,\alpha}$, see Eqn.~(\ref{HadTen-ell-3}),
into the expression for hadron tensor (\ref{HadTen1-4}).
After integration over $d^2\vec{\ell}_\perp$ and, then, after integration over  $d^2\vec{z}_\perp$
we get the following expression for the $\vec{\ell}_\perp$-term of the hadron tensor:
\begin{eqnarray}
\label{HadTen1-5}
&&\hspace{-0.3cm}{\cal W}^{(1)\, [\vec{\ell}_\perp]}_{\mu\nu}(g|A^\perp)=\int d\mu (k_i;x_1,y) \,
\, \bar\Phi^{[\gamma^-]} (k_2)
\text{tr}\Big[
\gamma_\mu  \gamma^- \gamma_\nu \gamma^+ \Big]\times
\nonumber\\
&&\hspace{-0.3cm}\int d^4\eta_1 \,
e^{-ik_1\eta_1}\times
\\
&&\hspace{-0.3cm}
\langle p_1| \bar\psi(\eta_1)\gamma^+
ig \int\limits_{-\infty^\perp}^{0^\perp} d\omega^\perp_\beta A^\perp_\beta(0^+, -\infty^-, \vec{\omega}^\perp)
\psi(0) | p_1\rangle \,.
\nonumber
\end{eqnarray}
As well as for the case of longitudinal gluons, if we now include all gluon emissions
from the antiquark going from the upper blob in Fig.~\ref{Fig-DY}, left panel,
we reproduce the corresponding $P$-exponential with the transverse gluons in
$\Phi^{(A)\,[\gamma^+]}_\alpha (k_1,\ell)$. Together with the result obtained above for the $A^+$-fields, we finally
have
\begin{eqnarray}
\label{me-Pexp-2}
&&\int d^4\eta_1 \, e^{-ik_1\cdot\eta_1}
\langle p_1 | \bar\psi(0^+,\eta_1^-, \vec{\bf 0}_\perp)\, \gamma^+ \times
\\
&&[0^+,-\infty^-,\vec{\bf 0}_\perp \,; 0^+,0^-,\vec{\bf 0}_\perp]_{A^+}
\times
\nonumber\\
&&[0^+,-\infty^-,-\vec{\bf\infty}_\perp \,; 0^+,-\infty^-,\vec{\bf 0}_\perp]_{A^\perp}
\psi(0) | p_1\rangle \, ,
\nonumber
\end{eqnarray}
where
\begin{eqnarray}
\label{Pexp-1-1}
&&[0^+,-\infty^-,-\vec{\bf\infty}_\perp \,; 0^+,-\infty^-,\vec{\bf 0}_\perp]_{A^\perp} =
\nonumber\\
&&\mathbb{P}{\rm exp}\Big\{ i g \int\limits^{-\infty^\perp}_{0^\perp}
 d\omega_\beta^\perp A_\beta^\perp(0^+,-\infty^-,\vec{{\bf\omega}}_\perp) \Big\}\, .
\end{eqnarray}
The transverse components of gluon fields, $A^\perp$, have the collinear twist
which equals to $1$. Therefore, the Wilson line in Eqn.~(\ref{Pexp-1-1}) represents
the infinite amount of the sub-dominant contributions. Within our frame,
it is enough to be limited by the collinear twist three contributions only.
In other words, we leave only the terms which include the first order of $A^\perp$.

\subsection*{The non-standard hadron tensor (direct process)}

The next step of our consideration is the contribution of the non-standard
diagram, depicted in Fig.~\ref{Fig-DY}, the right panel. The DY hadron tensor
receives the contribution from the non-standard diagram as (before factorization)
\begin{eqnarray}
\label{HadTen-NS-1}
&&{\cal W}^{(2)}_{\mu\nu}(g|A)=
\int d^4 k_1\, d^4 k_2 \, \delta^{(4)}(k_1+k_2-q)\times
\\
&&\text{tr}\big[
\gamma_\mu  {\cal F}(k_1) \gamma_\nu \bar\Phi(k_2)
\big]\, ,
\nonumber
\end{eqnarray}
where the function ${\cal F}(k_1)$ reads
\begin{eqnarray}
\label{PhiF2}
&&{\cal F}(k_1)= S(k_1) \gamma_\alpha \int d^4\eta_1\, e^{-ik_1\cdot\eta_1}\times
\\
&&\langle p_1| \bar\psi(\eta_1) \, gA_{\alpha}(0) \, \psi(0) | p_1\rangle \, .
\nonumber
\end{eqnarray}
Performing the above-described factorization procedure, the non-standard hadron tensor
takes the following form:
\begin{eqnarray}
\label{HadTen-NS-2}
&&{\cal W}^{(2)}_{\mu\nu}(g|A)=  \int dx_1 \, dy \,
\big[\delta(x_1-x_B) \delta(y-y_B)\big] \, \bar q(y) \times
\nonumber\\
&&\text{tr}\biggl[
\gamma_\mu \biggl( \int d^4 k_1\,
\delta(x_1p_1^+ - k_1^+) {\cal F}(k_1)\biggr) \gamma_\nu \hat p_2 \biggr]\equiv
\nonumber\\
&&\int dx_1 \, dy \,
\big[\delta(x_1-x_B) \delta(y-y_B)\big] \, \bar q(y)\,p_2^- \mathfrak{N}^+_{\mu\nu}(x_1)
 \, .
\end{eqnarray}
We now consider the integral over $k_1$ in (\ref{HadTen-NS-2}), we write
\begin{eqnarray}
\label{FacF2}
&&\mathfrak{N}^+_{\mu\nu}=\int d^4 k_1\, \delta(x_1p_1^+ - k_1^+)\times
\\
&&
\text{tr}\Big[ \gamma_\mu \frac{k_1^+\gamma^- + k_1^-\gamma^+ - \vec{k}_{1\perp}\vec{\gamma}_\perp}
{2k_1^+k_1^- - \vec{k}_{1\perp}^2 +i\epsilon} \gamma_\alpha \gamma^- \gamma_\nu \gamma^+
\Big]\times
\nonumber\\
&&
\int d^4\eta_1\, e^{-ik_1\cdot\eta_1}
\langle p_1 | \bar\psi(\eta_1) \gamma^+  gA_{\alpha}(0) \psi(0) | p_1\rangle.
\nonumber
\end{eqnarray}
Technically, derivation of the longitudinal Wilson line for this case differs from the
derivation we implemented for the standard hadron tensor.
We notice that for the non-standard hadron tensor the quark propagator has been included
in the soft part.

Let us consider
the first term, $k_1^+\gamma^-$, in the quark propagator, see Eqn.~(\ref{FacF2}).
Thanks for the $\gamma$-structure, this term singles out the $A^-$-field in the corresponding
correlator. Moreover, the Fourier image of the quark-gluon correlator can be presented
in the equivalent form as
\begin{eqnarray}
\label{FI-1}
&&\int d^4\eta_1\, e^{-ik_1\cdot\eta_1-ik_1 z}
\langle p_1 | \bar\psi(\eta_1) \gamma^+ \times
\\
&&g
\frac{\partial}{\partial z^+} \int\limits_{-\infty^+}^{z^+} d\omega^+ A^-(\omega^+,0^-,\vec{\bf 0}_\perp)
\Bigg|_{z=0}
 \psi(0) | p_1\rangle,
\nonumber
\end{eqnarray}
where the derivative with respect to $z^+$ can be shifted to the exponential function
$e^{-ik_1^- z^+}$. As a result, we have
\begin{eqnarray}
\label{FI-2}
&&i k_1^- \int d^4\eta_1\, e^{-ik_1\cdot\eta_1}\times
\\
&&\langle p_1 | \bar\psi(\eta_1) \gamma^+
g
\int\limits_{-\infty^+}^{0^+} d\omega^+ A^-(\omega^+,0^-,\vec{\bf 0}_\perp)
 \psi(0) | p_1\rangle.
\nonumber
\end{eqnarray}
Using Eqn.~(\ref{FI-2}), the tensor $\mathfrak{N}_{\mu\nu}$ takes the form of
($\vec{k}_{1\perp}^{\,2}\approx 0$)
\begin{eqnarray}
\label{FacF3}
&&\mathfrak{N}^+_{\mu\nu}=\int d^4 k_1\, \delta(x_1p_1^+ - k_1^+)
\,\text{tr}\big[ \gamma_\mu \gamma^- \gamma_\nu \gamma^+
\big]\times
\nonumber\\
&&
\int d^4\eta_1\, e^{-ik_1\cdot\eta_1}\times
\\
&&
\langle p_1 | \bar\psi(\eta_1) \gamma^+
ig\int\limits_{-\infty^+}^{0^+} d\omega^+ A^-(\omega^+,0^-,\vec{\bf 0}_\perp) \psi(0) | p_1\rangle.
\nonumber
\end{eqnarray}
Thus, the first term finally contributes to the non-standard part of the hadron tensor as
\begin{eqnarray}
\label{HadTen-NS-3}
&&\hspace{-.8cm}\overline{\cal W}^{(0)}_{\mu\nu}(A^-)=  \int dx_1 \, dy \,
\big[\delta(x_1-x_B) \delta(y-y_B)\big] \, \bar q(y) \times
\nonumber\\
&&\hspace{-.8cm}\int d^4 k_1\, \delta(x_1p_1^+ - k_1^+)
\,\text{tr}\big[ \gamma_\mu \gamma^- \gamma_\nu \gamma^+
\big]
\int d^4\eta_1\, e^{-ik_1\cdot\eta_1}\times
\\
&&\hspace{-.8cm}
\langle p_1 | \bar\psi(\eta_1) \gamma^+
[-\infty^+,0^-,\vec{\bf 0}_\perp \,; 0^+,0^-,\vec{\bf 0}_\perp]_{A^-}
\psi(0) | p_1\rangle\,.
\nonumber
\end{eqnarray}
The exponentiation of $A^-$ has been presented in Appendix~\ref{Exp-Aminus:App:A}.

Despite the minus component, $A^-$, has formally the collinear twist $2$
(the so-called sub-sub-dominant component),
the Wilson line with $A^-$ in Eqn.~(\ref{HadTen-NS-3})
will play the substantial role for the residual gauge fixing, see discussion in the next section.

\begin{figure}[t]
\centerline{\includegraphics[width=0.45\textwidth]{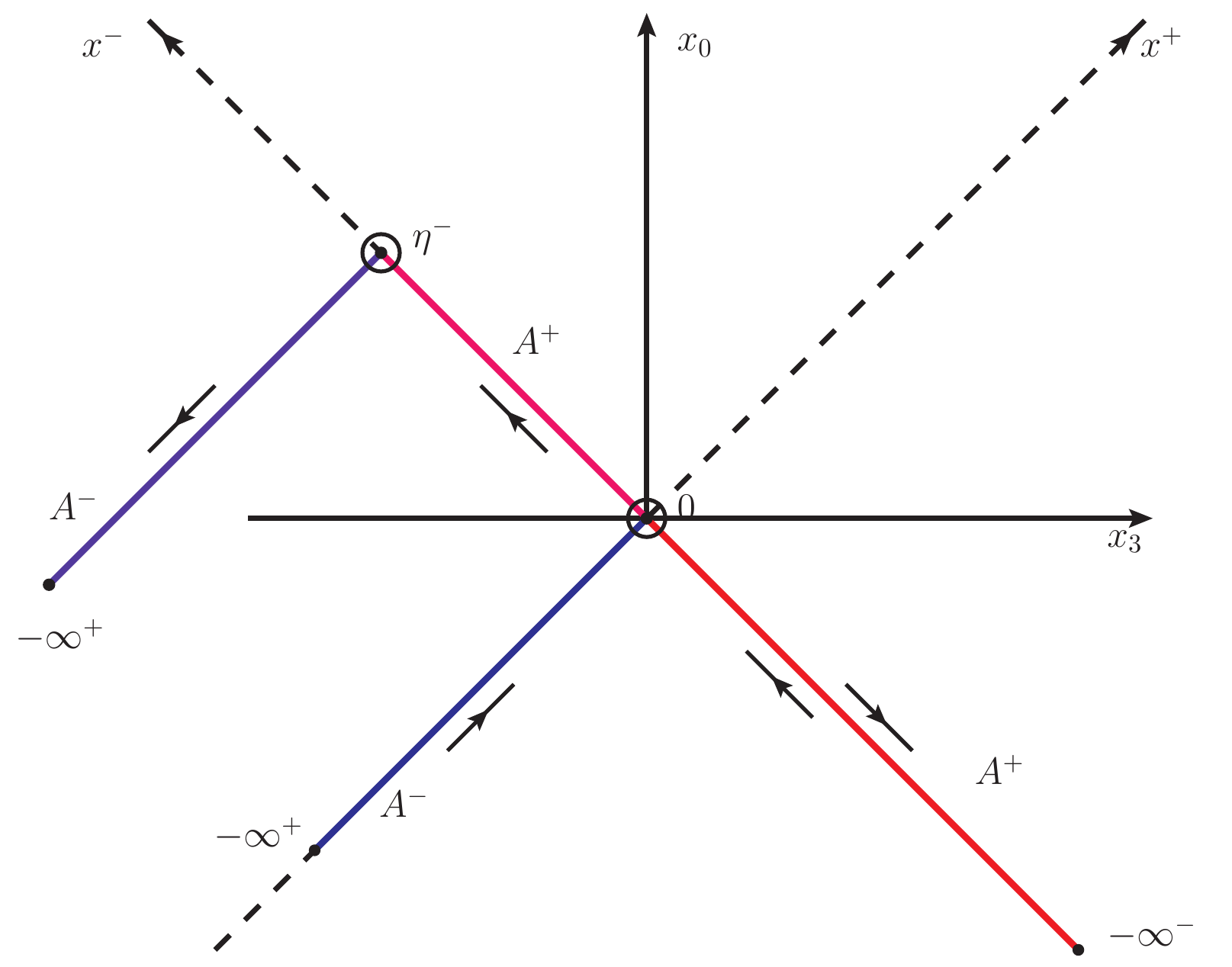}}
  \caption{The longitudinal Wilson lines related to the standard (red lines, the
  exponentials with $A^+$) and
  non-standard (blue lines, the exponentials with $A^-$) Drell-Yan hadron tensor.
  The circles single out the interception points which the continuity conditions
  are defined for.}
\label{Fig-WL}
\end{figure}

To conclude the section, we restore all the longitudinal Wilson lines which emanate from
both the standard and non-standard hadron tensors, see Fig.~\ref{Fig-WL}.

\section{Contour Gauge: Elimination of longitudinal Wilson lines}

The axial gauge $A^+=0$ (as well as the Fock-Schwinger gauges)
is in fact a particular case of the most general non-local contour gauge
determined by a Wilson line with a fixed path. Indeed, the straightforward line
in the Wilson exponential which
connects $\pm\infty$ with $x$ gives us the axial gauge, while
the straightforward line connecting $x_0$ with $x$ leads to the Fock-Schwinger gauge.
Notice that two different contour gauges can correspond to the same
local axial gauge. Meanwhile, to distinguish different contour gauge
is very crucial to fix the prescriptions in the gluon poles \cite{AT-GP}.

In the past, the contour gauge was very popular subject of intense studies
(see, for example, \cite{ContourG}).
One of the advantages of using the contour gauge
is that the quantum gauge theory becomes free from the Gribov ambiguities.
On the other hand, the contour gauge gives the simplest way to fix the
gauge including the residual gauge freedom.
In contrast to the usual axial gauge, in the contour gauge we first  fix an arbitrary
point $(x_0, \textbf{g}(x_0))$
in the fiber. Then, we define two directions: one of them in the base, the other in
the fiber. The direction in the base $\mathbb{R}^4$ is nothing else than the tangent vector of a curve which
goes through the given point $x_0$. The fiber direction  can be uniquely determined as the
tangent subspace related to the parallel transport. Finally, we are able to
define uniquely the point in the fiber bundle.

We continue to work with the Drell-Yan hadron tensor derived in \cite{AT-GP}.
As shown the standard (direct and mirror) diagrams lead to the following
Wilson lines in the quark-antiquark nonlocal operator which forms the hadron tensor, see Fig.~\ref{Fig-WL}:
\begin{eqnarray}
\label{WL1-mir}
&&[0^+,\eta^-,\vec{\bf 0}_\perp; \, 0^+,-\infty^-,\vec{\bf 0}_\perp]_{A^+},\quad \text{and}
\\
\label{WL1-dir}
&&[0^+,-\infty^-,\vec{\bf 0}_\perp; \, 0^+,0^-,\vec{\bf 0}_\perp]_{A^+},
\end{eqnarray}
{\it i.e.} the gauge invariant quark string operator takes the form of
\begin{eqnarray}
\label{OP1}
&&\bar\psi(0^+,\eta^-,\vec{\bf 0}_\perp) [0^+,\eta^-,\vec{\bf 0}_\perp; \, 0^+,-\infty^-,\vec{\bf 0}_\perp]_{A^+}
\Gamma\times
\nonumber\\
&&[0^+,-\infty^-,\vec{\bf 0}_\perp; \, 0^+,0^-,\vec{\bf 0}_\perp]_{A^+} \psi(0^+,0^-,\vec{\bf 0}_\perp).
\end{eqnarray}
Here $\Gamma$ implies a relevant combination of $\gamma$-matrices.
The Wilson line (\ref{WL1-mir}) is a result of summation in the mirror diagram and
the Wilson line (\ref{WL1-dir}) appears in the direct diagram.

The sum of direct and mirror diagram contributions takes place
if we study the spin-average DY hadron tensor. While, for the single transverse
spin asymmetry, we deal individually with only the direct (or mirror) diagram contribution
because the direct and mirror diagrams differ in sign to construct the corresponding SSA.
For our further considerations in the context of contour gauge, it is not so
crucial what kind of hadron tensors we work with.

The non-standard (direct and mirror) diagrams give us the contributions with
the Wilson lines
\begin{eqnarray}
\label{WL2}
&&[-\infty^+,\eta^-,\vec{\bf 0}_\perp; \, 0^+,\eta^-,\vec{\bf 0}_\perp]_{A^-},\quad \text{and}
\nonumber\\
&&[0^+,0^-,\vec{\bf 0}_\perp; \, -\infty^+,0^-,\vec{\bf 0}_\perp]_{A^-},
\end{eqnarray}
and, therefore, we have the string operator
\begin{eqnarray}
\label{OP2}
&&\bar\psi(0^+,\eta^-,\vec{\bf 0}_\perp) [-\infty^+,\eta^-,\vec{\bf 0}_\perp; \, 0^+,\eta^-,\vec{\bf 0}_\perp]_{A^-}
\Gamma\times
\nonumber\\
&&[0^+,0^-,\vec{\bf 0}_\perp; \, -\infty^+,0^-,\vec{\bf 0}_\perp]_{A^-} \psi(0^+,0^-,\vec{\bf 0}_\perp).
\end{eqnarray}

According to the contour gauge conception,
we eliminate all the Wilson lines with the longitudinal (unphysical) gluon fields $A^+$ and $A^-$.
We note that the ideologically similar approach can be found in \cite{Belitsky:2002sm}.

We begin with the Wilson lines shown in Eqns.~(\ref{WL1-mir})-(\ref{WL1-dir}),
we write the following gauge fixing conditions:
\begin{eqnarray}
\label{CG-1}
&&[0^+,\eta^-,\vec{\bf 0}_\perp; \, 0^+,-\infty^-,\vec{\bf 0}_\perp]_{A^+}=\mathds{1},
\nonumber\\
&&[0^+,-\infty^-,\vec{\bf 0}_\perp; \, 0^+,0^-,\vec{\bf 0}_\perp]_{A^+}=\mathds{1}
\end{eqnarray}
explicit solutions of which read
\begin{eqnarray}
\label{Aplus-1}
&&A^+(0^+,\mathds{L}_{0^-,-\infty^-},\vec{\bf 0}_\perp)=0,
\\
\label{Aplus-2}
&&A^+(0^+,\mathds{L}_{-\infty^-,\eta^-},\vec{\bf 0}_\perp) = 0.
\end{eqnarray}
Here $\mathds{L}_{x,y}$ denotes the straightforward line in the Minkowski space connecting point $x$ with point $y$.
In the contour gauge (\ref{CG-1})-(\ref{Aplus-2}), the remaining gluon field components can be represented as
(with $\mu=-,\perp$)
\begin{eqnarray}
\label{InRep-1}
&&A^{\mu}_G(0^+,x^-,\vec{\bf 0}_\perp)\Big|_{c.g.(\ref{CG-1})-(\ref{Aplus-2})}
= \int\limits_{-\infty^-}^{x^-} dz_\alpha \frac{\partial z_\beta}{\partial x_\mu}
G^{\alpha\beta}(z|A^{\mu}_{c.g.})
\nonumber\\
&&
=\tilde{n}^-\int\limits_{0}^{\infty} ds e^{-\epsilon s} G^{+\mu}(x^-- \tilde{n}^-s|A^{\mu}_{c.g.})
\end{eqnarray}
with the boundary condition
\begin{eqnarray}
\label{bc}
A^{\mu}_{b.c.}(0^+,x^--\tilde{n}^-\infty,\vec{\bf 0}_\perp)\Big|_{c.g.(\ref{CG-1})-(\ref{Aplus-2})}=0.
\end{eqnarray}
In Eqn.~(\ref{InRep-1}), we use the parametrization of $\mathds{L}_{-\infty^-,x^-}$ as
\begin{eqnarray}
\label{ParL}
&&z(s)=\big(0^+,x^--\tilde{n}^-\lim\limits_{\epsilon\to 0}\frac{1-e^{-\epsilon s}}{\epsilon},\vec{\bf 0}_\perp\big),
\\
&&dz_{\alpha}\Big|^{-\infty}_{x}=\tilde{n}_\alpha ds e^{-\epsilon s} \Big|_{0}^{\infty}.
\nonumber
\end{eqnarray}

We now dwell on the gauge conditions for $A^-$ gluon component.
We put the Wilson lines (\ref{WL2}) to be equal to $1$ too, {\it i.e.}
\begin{eqnarray}
\label{CG-2}
&&[-\infty^+,\eta^-,\vec{\bf 0}_\perp; \, 0^+,\eta^-,\vec{\bf 0}_\perp]_{A^-}=\mathds{1},
\nonumber\\
&&[0^+,0^-,\vec{\bf 0}_\perp; \, -\infty^+,0^-,\vec{\bf 0}_\perp]_{A^-}=\mathds{1}.
\end{eqnarray}
These conditions yield
\begin{eqnarray}
\label{Aminus-1}
&&A^-(\mathds{L}_{0^+,-\infty^+},\eta^-,\vec{\bf 0}_\perp)=0,
\\
\label{Aminus-2}
&&A^-(\mathds{L}_{-\infty^+,0^+},0^-,\vec{\bf 0}_\perp) = 0.
\end{eqnarray}
As above, in the contour gauge (\ref{CG-2})-(\ref{Aminus-2}), the remaining
gluon fields have the integral representations which read
(here $\mu=+,\perp$)
\begin{eqnarray}
\label{InRep-2}
&&A^{\mu}_G(x^+,\eta^-,\vec{\bf 0}_\perp)\Big|_{c.g.(\ref{CG-2})-(\ref{Aminus-2})}
= \int\limits^{-\infty^+}_{x^+} dz_\alpha \frac{\partial z_\beta}{\partial x_\mu}
G^{\alpha\beta}(z|A^{\mu}_{c.g.})
\nonumber\\
&&=-\breve n^+ \int\limits_{0}^{\infty} dt e^{-\epsilon t}
G^{-\mu}\big(x^+ - \breve n^+ t|A^{\mu}_{c.g.}\big)
\end{eqnarray}
with the boundary condition
\begin{eqnarray}
\label{bc-1}
A^{\mu}_{b.c.}(x^+ -\breve n^+ \infty,\eta^-, \vec{\bf 0}_\perp)
\Big|_{c.g.(\ref{CG-2})-(\ref{Aminus-2})}=0.
\end{eqnarray}
In Eqn.~(\ref{InRep-2}) the path parametrization of $\mathds{L}_{x,-\infty}$ is given by
\begin{eqnarray}
\label{ParL-1}
&&z(s)=\big(x^+ -\breve n^+\lim\limits_{\epsilon\to 0}\frac{1-e^{-\epsilon t}}{\epsilon},\eta^-,\vec{\bf 0}_\perp\big),
\\
&&dz_{\alpha}\Big|^{-\infty}_{x}=-\breve n^+_{\alpha} dt e^{-\epsilon t} \Big|_{0}^{\infty}.
\nonumber
\end{eqnarray}

Further, the gluon field $A^{-}_G$ of Eqn.~(\ref{InRep-1}) has to be
compatible with the gluon field $A^-$ of Eqn.~(\ref{Aminus-1}).
Also, the same inference has to be valid for the gluon fields $A^{+}_G$ of Eqn.~(\ref{InRep-2})
and $A^+$ of Eqn.~(\ref{Aplus-2}).
We thus require the analytical continuity for these gluon fields at the interception points,
see Fig.~\ref{Fig-WL},
and we finally arrive at the following conditions (here we omit the subscript $G$)
\begin{eqnarray}
\label{Gfix-1}
&&A^+(0^+,x^-=\eta^-,\vec{\bf 0}_\perp) = A^+(x^+=0^+,\eta^-,\vec{\bf 0}_\perp) =0,
\nonumber\\
&&A^-(x^+=0^+,\eta^-,\vec{\bf 0}_\perp) = A^-(0^+,x^-=\eta^-,\vec{\bf 0}_\perp) =0,
\nonumber\\
\end{eqnarray}
respectively. Having used these conditions, we stay with the
physical gluon fields $A^\perp$ only.

\section{Gluon Propagator}

We now go over to consideration of the gluon propagator.
In the case of local axial gauge $A^+=0$, the gluon propagator
is still not a well-defined object because of the spurious singularity
related to the residual gauge transformations.
In other words, the axial gauge cannot fix completely the unique element of each
orbit defined on the gauge group. In Appendix~\ref{G-Rsymmetry:App:B},
we present the handbook material regarding the gauge and residual gauge fixing.
It is clear that if, in the local axial gauge $A^+=0$, we fix the residual gauge
by requiring $\theta^a_0(k^-,\vec{\bf k}_\perp)=0$ (see, Eqns.~(\ref{ReAx-con-p-1})-(\ref{ReAx-con-reg}))
we immediately get that $A^-=0$ as well. The same inference can be reached by the simplest way
if we use the contour gauge conception (see, Eqn.~(\ref{Gfix-1})).
Notice that the maximal gauge fixing which is based on the contour gauge conception
does not relate technically to the problem of finding the inverse kinematical operator
(see, Eqns.~(\ref{Kop})-(\ref{Cont-Eqn})). The contour gauge approach is, therefore,
an alternative method of gauge fixing compared to the ``classical'' approaches based on the
corresponding effective Lagrangian (see, for example, \cite{Chirilli:2015fza}).

So, we perform our calculation in the contour gauge defined by Eqns.~(\ref{CG-1}) and/or (\ref{CG-2})
together with the conditions of Eqn.~(\ref{Gfix-1}) where the only physical gluons are presented.
In the framework of collinear factorization under our consideration,
the gluon momentum has the plus dominant components.

Having used the Wilson lines from the standard and non-standard diagrams,
we calculate the gluon propagator which reads
\begin{eqnarray}
\label{GP-1}
\langle 0| T A^{\mu}_\perp(0^+,x^-, \vec{\bf 0}_\perp)
A^{\nu}_\perp(0^+,0^-, \vec{\bf 0}_\perp)|0\rangle = D^{\mu\nu}_\perp(x^-).
\end{eqnarray}
Using the integral representation (\ref{InRep-1}), the gluon propagator takes the form of
\begin{eqnarray}
\label{GP-2}
&&D^{\mu\nu}_\perp(x^-)=
n_\alpha n_\beta \int\limits_0^\infty ds ds^\prime e^{-\epsilon s - \epsilon s^\prime}
\times
\nonumber\\
&&
\langle 0| T G^{\mu\alpha}(x^--\tilde n^-s) G^{\nu\beta}(0^--\tilde n^-s^\prime)|0\rangle=
\nonumber\\
&&\int(d^4 \ell) e^{-i\ell^+x^-} \frac{1}{\ell^2+i0}
\frac{(\ell^+)^2 d^{\mu\nu}_\perp(\ell)}{(\ell^++i\epsilon)(\ell^+-i\epsilon)}.
\end{eqnarray}
In Eqn.~(\ref{GP-2}), we have explicitly performed the integration over $ds(ds^\prime)$:
\begin{eqnarray}
\label{s-int}
\int\limits_0^\infty ds e^{\pm is(\ell^+ \pm i\epsilon)} = \frac{\pm i}{\ell^+ \pm i\epsilon}
\end{eqnarray}
which emanates from the path parametrization.
It is worth to emphasize the gluon pole prescription can be traced from this
kind of integrations (see, \cite{AT-GP}).
The transverse tensor $d^{\mu\nu}_\perp$ has been constructed as
\begin{eqnarray}
\label{d-tensor}
d^{\mu\nu}_\perp(\ell)= g^{\mu\nu}- \frac{\ell^{\mu,+} n^\nu + \ell^{\nu,+} n^\mu}{[\ell^+]_{\rm reg}}
\end{eqnarray}
where the spurious singularity $[\ell^+]_{\rm reg}$ has to be regularized.

We consider the combination
\begin{eqnarray}
\label{com}
\frac{(\ell^+)^2}{(\ell^++i\epsilon)(\ell^+-i\epsilon)}
d^{\mu\nu}_\perp(\ell).
\end{eqnarray}
The first term of Eqn.~(\ref{com}) includes the combination
\begin{eqnarray}
g^{\mu\nu}\ell^+ \frac{\ell^+}{(\ell^++i\epsilon)(\ell^+-i\epsilon)}
\end{eqnarray}
which has to be treated only as
\begin{eqnarray}
\label{g-met}
g^{\mu\nu}\frac{\ell^+}{2}\Big( \frac{1}{\ell^++i\epsilon} + \frac{1}{\ell^+-i\epsilon} \Big)
=g^{\mu\nu}\ell^+\frac{\mathcal{P}}{\ell^+}=g^{\mu\nu}.
\end{eqnarray}

On the other hand, for $x^- >0$ (see, the momentum integral (\ref{GP-2})), the integration contour
has to be closed in the lower semi-plane, $\Im\text{m} \ell^+ < 0$.
Hence, for the $g_{\mu\nu}$-term, we obtain the integrand
\begin{eqnarray}
\label{int-g}
g^{\mu\nu} \frac{\ell^+}{\ell^++i\epsilon}
\end{eqnarray}
where the denominator $\ell^+-i\epsilon$ has been cancelled by one of $\ell^+$ in the numerator.
It is clear that the remaining combination in Eqn.~(\ref{int-g}) yields $g_{\mu\nu}$ (cf. Eqn.~(\ref{g-met})).

Regarding the second term of Eqn.~(\ref{com}), we propose two ways of reasoning.

\noindent
{\it The first way:} We don't specify explicitly the tensor structure of this term.
The second term of Eqn.~(\ref{com}) can be written in the following form
(here the momentum flux direction is not fixed):
\begin{eqnarray}
\label{sec-term-1}
&&\frac{(\ell^+)^2}{(\ell^++i\epsilon)(\ell^+-i\epsilon)} \frac{L^{\mu\nu}(\ell,n)}{[\ell^+]_{\rm reg}}=
\nonumber\\
&&\ell^+\frac{\mathcal{P}}{\ell^+} \frac{L^{\mu\nu}(\ell,n)}{[\ell^+]_{\rm reg}},
\end{eqnarray}
where we use
\begin{eqnarray}
\frac{\mathcal{P}}{\ell^+}=\frac{\ell^+}{(\ell^+  + i\epsilon)(\ell^+ - i\epsilon)}.
\end{eqnarray}

To well-define the product of two generalized functions the pole $1/[\ell^+]_{\rm reg}$ must be treated only as
\begin{eqnarray}
\label{Reg-pv}
\frac{1}{[\ell^+]_{\rm reg}}=\frac{\mathcal{P}}{\ell^+}.
\end{eqnarray}
Indeed, we have
\begin{eqnarray}
\frac{\mathcal{P}}{\ell^+} \ell^+ \frac{\mathcal{P}}{\ell^+} = \frac{\mathcal{P}}{\ell^+}.
\end{eqnarray}
On the other hand, if we let $1/[\ell^+]_{\rm reg}$ be equal to $1/(\ell^+\pm i\epsilon)$,
we will face on the wrong-defined product of two generalized functions \cite{Vladimirov}:
\begin{eqnarray}
&&\frac{\mathcal{P}}{\ell^+} \ell^+ \frac{1}{\ell^+\pm i\epsilon} =
\frac{\mathcal{P}}{\ell^+} \ell^+ \Big( \frac{\mathcal{P}}{\ell^+} \mp i\pi \delta(\ell^+)\Big)
\nonumber\\
&&\Longrightarrow \frac{\mathcal{P}}{\ell^+} \ell^+ \delta(\ell^+)\quad \text{-- wrong-defined product}.
\end{eqnarray}

\noindent
{\it The second way:} We take into account that the tensor structure includes
the plus component of the gluon momentum. Hence, the second term of of Eqn.~(\ref{com})
reads
\begin{eqnarray}
\label{sec-term-2}
\frac{(\ell^+)^2}{(\ell^++i\epsilon)(\ell^+-i\epsilon)} \frac{\ell^{\mu,+} n^\nu + \ell^{\nu,+} n^\mu}{[\ell^+]_{\rm reg}}.
\end{eqnarray}
Here, as shown above, for the first factor, we can again use that
\begin{eqnarray}
\label{sec-term-2-1}
\frac{(\ell^+)^2}{(\ell^++i\epsilon)(\ell^+-i\epsilon)}=\ell^+\frac{\mathcal{P}}{\ell^+}=1
\end{eqnarray}
and, for the second factor, we have
\begin{eqnarray}
\label{sec-term-2-2}
\frac{\ell^{\mu,+} n^\nu + \ell^{\nu,+} n^\mu}{[\ell^+]_{\rm reg}} =
\frac{\ell^+}{[\ell^+]_{\rm reg}} \big( n^{*\,\mu} n^\nu + n^{*\,\nu} n^\mu\big).
\end{eqnarray}
Based on this expression, it is clear that the only possibility
is to define $1/[\ell^+]_{\rm reg}$ through the principle value, see Eqn.~(\ref{Reg-pv}).

Thus, in the contour gauge generated by both the standard and non-standard diagrams,
the gluon propagator reads
\begin{eqnarray}
\label{GP-f1}
&&\hspace{-0.5cm}D^{\mu\nu}_\perp(x^-)=
\\
&&\hspace{-0.5cm}\int(d^4 \ell) e^{-i\ell^+x^-} \frac{1}{\ell^2+i0}
\Big\{ g^{\mu\nu} - \frac{\mathcal{P}}{\ell^+}\big(\ell^{\mu,+} n^\nu + \ell^{\nu,+} n^\mu\big) \Big\}
\nonumber
\end{eqnarray}
or, using Eqn.~(\ref{sec-term-2-1}), we obtain
\begin{eqnarray}
\label{GP-f1-2}
D^{\mu\nu}_\perp(x^-)=
\int(d^4 \ell) e^{-i\ell^+x^-} \frac{g^{\mu\nu}_\perp}{\ell^2+i0},
\end{eqnarray}
where $g^{\mu\nu}_\perp = g^{\mu\nu} -  n^{*\,\mu} n^\nu - n^{*\,\nu} n^\mu$.

We notice that the gluon propagator presented in Eqn.~(\ref{GP-f1-2}) takes place for the
very specific case of the polarized DY hadron tensor under our consideration.
In the case of deep-inelastic scattering process, where the corresponding Wilson lines are different,
the gluon propagator derived in the contour gauge frame has the form similar to Eqn.~(\ref{GP-f3}), see below.
We also stress that, in Eqns.~(\ref{GP-f1}) and (\ref{GP-f1-2}), the gluon momentum flux is not important and is not specified.

We now consider a particular case wherein only the standard diagram exists.
For example, this can be achieved if we neglect the higher twist correlators
$\langle \bar\psi A^- \psi\rangle$
which appear in the non-standard diagram.
Moreover, the gluon field co-ordinates are not
necessarily on the minus direction and
the gluon momentum flux is fixed
in the positive direction from the $\nu$-vertex to $\mu$-vertex.
In this case, the gluon propagator reads
\begin{eqnarray}
\label{GP-f2}
&&D^{\mu\nu}(x)\Big|^{\text{stand. dia.}}_{\text{fixed flux}}=
\int(d^4 \ell) e^{-i\ell x} \frac{1}{\ell^2+i0}\times
\\
&&\Big\{ g^{\mu\nu} - \frac{\mathcal{P}}{\ell^+}\bigg(\ell^{\mu} n^\nu\theta(\ell^+)
+ \ell^{\nu} n^\mu \theta(-\ell^+)\bigg) \Big\}
\nonumber
\end{eqnarray}
where the corresponding $\theta$-functions specify the momentum flux.
Using the Cauchy theorem in Eqn.~(\ref{GP-f2}), we finally arrive at
\begin{eqnarray}
\label{GP-f3}
&&D^{\mu\nu}(x)\Big|^{\text{stand. dia.}}_{\text{fixed flux}}=\int(d^4 \ell)  \frac{e^{-i\ell x}}{\ell^2+i0}\times
\\
&&\Big\{ g^{\mu\nu} - \frac{\ell^{\mu} n^\nu}{\ell^+- i\epsilon} -
\frac{\ell^{\nu} n^\mu}{\ell^+ + i\epsilon} \Big\}
\nonumber
\end{eqnarray}
which coincides with the results in \cite{Belitsky:2002sm}, \cite{Chirilli:2015fza}.
This expression is sensitive to the definition of the positive (negative)
flux direction (see, Eqn.~(\ref{GP-f2})).
Hence, the symmetry over $\mu\leftrightarrow\nu$ takes place only together with
the simultaneous replacement $\ell^+ \leftrightarrow -\ell^+$ in the second and third terms of
Eqn.~(\ref{GP-f3}).

\section{Conclusions and discussions}

In the contour gauge, from the technical viewpoint, the maximal gauge fixing are not associated with
the problem of finding the inverse kinematical operator. Hence,
the contour gauge approach has to be considered as
the alternative method of gauge fixing in comparison with the ``classical'' approaches based on the
corresponding effective Lagrangians.
It is necessary to stress that the contour gauge contains the important and unique additional information (needed to fix
the prescription in the gluon poles) which is invisible in the case of usual (local) axial gauge.
From this point of view, before we discard the terms with $A^+$, we have to determine
the relevant fixed path in the corresponding Wilson line with $A^+$
which finally leads to the certain prescriptions in the gluon poles.
Moreover, the corresponding Wilson line with $A^-$ in the non-standard diagram, which contributes to the
polarized DY hadron tensor, prompts the way of residual gauge fixing.

We thus advocate the preponderance of the contour gauge use which allows to fix completely
the gauge freedom by the most illustrative and simplest way. We demonstrate that
the non-standard diagram plays the important role in forming of the relevant contour
in the corresponding Wilson line. Hence, from the viewpoint of contour gauge, there is no
way to neglect the additional non-standard diagram.

\section*{Acknowledgments}

We thank A.V.~Efremov, A.~Prokudin and L.~Szymanowski for useful discussions.
The work by I.V.A. was partially supported by the Heisenberg-Landau Program of the
German Research Foundation (DFG). I.V.A. also thanks the Department of Physics at the
University of Antwerpen for warm hospitality.

\appendix
\renewcommand{\theequation}{\Alph{section}.\arabic{equation}}

\section{Exponentiation of component $A^-$}
\label{Exp-Aminus:App:A}

In this Appendix, we demonstrate the method of the $A^-$ component exponentiation.
In fact, there are several methods how to exponentiate the gluon fields,
see e.g. \cite{Gross:1971wn, Balitsky:1987bk, Radyushkin:1983mj}. Here, we present an alternative
frame-independent and most efficient method mainly based on the approach described in \cite{BogoShir}, see \S 46.

\subsection*{Some conventional notations}

Before going further, we remind several conventions regarding how the gauge transformations match
the Wilson lines. Taking, for the sake of simplicity, the Abelian gauge theory
(in the case of interest the distinction between Abelian and non-Abelian groups is irrelevant),
let us assume that the fermion and gauge fields are transformed as
\begin{eqnarray}
\label{Wl-GT-1}
&&\psi^{\omega}(x)=e^{\pm i\theta(x)} \psi(x),
\\
\label{Wl-GT-1-2}
&&A_{\mu}^\omega(x)=A_\mu(x) \pm \partial_\mu\theta(x)
\end{eqnarray}
where  $\omega$ stands here for the gauge transformation.
Generally speaking, the signs at the gauge function $\theta$ in Eqns.~(\ref{Wl-GT-1}) and (\ref{Wl-GT-1-2}) are
conventional. If we fix the transformations as in Eqns.~(\ref{Wl-GT-1}) and (\ref{Wl-GT-1-2}), {\it i.e.}
the same sings in both expressions,
then we can readily see that the covariant derivative and
the {\it gauge-invariant} fermion string operator become
\begin{eqnarray}
\label{Wl-GT-2}
&&i{\cal D}_{\mu} = i\partial_\mu + g A_\mu(x),
\nonumber\\
&&\mathbb{O}^{\text{g.-inv.}}(x,y)=\bar\psi(y) [y ;\, x]_{A} \psi(x),
\end{eqnarray}
where the Wilson line is given by
\begin{eqnarray}
\label{Def-WL}
&&[y \,;\, x]_{A} \stackrel{{\rm def}}{=}
\mathbb{P}{\rm exp}\Big\{+ i g \int\limits^{y}_{x}
 dz_\mu \, A^\mu(z) \Big\}=
\\
&&\lim_{N\to\infty} [y \,;\, x_N]_{A}\,[x_N \,;\, x_{N-1}]_{A}...[x_1 \,;\, x]_{A}=
\nonumber\\
&&\lim_{N\to\infty} \big[1+igA(x_N)\cdot (y-x_N)\big]...\big[1+igA(x)\cdot (x_1-x)\big].
\nonumber
\end{eqnarray}
In Eqn.~(\ref{Def-WL}), the starting point $x$ and final point $y$ are connected by the certain path $\mathbb{P}\in \mathds{R}^4$
which allows the arrangement by pounding $\big\{ x_N \big\}^{y}_{x}$.

However, if the signs in both fermion and gauge boson transformations differ from each other, {\it i.e.}
\begin{eqnarray}
\label{Wl-GT-1-d}
&&\psi^{\omega}(x)=e^{\pm i\theta(x)} \psi(x),
\\
\label{Wl-GT-1-2-d}
&&A_{\mu}^\omega(x)=A_\mu(x) \mp \partial_\mu\theta(x),
\end{eqnarray}
the covariant derivative takes the form of
\begin{eqnarray}
\label{Wl-GT-2-d}
i{\cal D}_{\mu} = i\partial_\mu - g A_\mu(x),
\end{eqnarray}
while the {\it gauge-invariant} fermion string operator, in this case, reads
(see, for example, \cite{Belitsky:2002sm})
\begin{eqnarray}
\label{Wl-GT-2-2-d}
\mathbb{O}^{\text{g.-inv.}}(x,y)=\bar\psi(y) [x \,;\, y]_{A} \psi(x)
\end{eqnarray}
or
\begin{eqnarray}
\label{Wl-GT-2-2-d-2}
\mathbb{O}^{\text{g.-inv.}}(x,y)=\bar\psi(y) [y \,;\, x]^{-1}_{A} \psi(x)
\end{eqnarray}
with the Wilson line defined as in Eqn.~(\ref{Def-WL}).

In our paper, we adhere the conventions as in Eqn.~(\ref{Wl-GT-2}).

\subsection*{Description of the method}

We begin with the most illustrative subject which is the Green function in the external field.
The gluon radiation from the proper spinor line as shown in Fig.~\ref{Fig-DY}, the right panel,
is actually relevant to the Green function in the external field.

Consider the differential equation for the Green function
\begin{eqnarray}
\label{GF-1}
\big[i\widehat{\partial} + g \widehat{A}(x) \big] {\mathrm S}(x,y) = -\delta^{(4)}(x-y),
\end{eqnarray}
where the wide hat denotes the convolution with $\gamma$-matrices as $\widehat{A}=\gamma\cdot A$ etc.

We emphasize that the Green function defined by Eqn.~(\ref{GF-1}) is {\it not} gauge-invariant subject
(see, for example, \cite{Gross:1971wn, Balitsky:1987bk}).
As one can see below, namely the {\it gauge-noninvariant} Green function ensures the appearance of
the {\it gauge-invariant} fermion string operator in the corresponding hadron matrix element.

For the sake of simplicity and without the loss of generality, we assume that
$\partial^\mu=(0^+,\partial^-, \vec{\bf 0}_\perp)$, $A^\mu=(0^+,A^-, \vec{\bf 0}_\perp)$
and we, therefore, study the tensor combination as
${\mathrm S}^{[\gamma^+]}(x,y)\stackrel{{\rm def}}{=}\gamma^+{\mathrm S}(x,y)$. That is,
instead Eqn.~(\ref{GF-1}) we deal with the following differential equation
\begin{eqnarray}
\label{GF-2}
\big[i\partial^- + g A^-(x) \big] {\mathrm S}^{[\gamma^+]}(x,y) = -\delta^{(4)}(x-y).
\end{eqnarray}
Hence, in the operator forms, the Green function takes the form of
\begin{eqnarray}
\label{GF-3}
{\mathrm S}^{[\gamma^+]}(x,y) = -\frac{1}{\big[i\hat\partial^- + g \hat A^-(x) \big]}\delta^{(4)}(x-y)
\end{eqnarray}
where the small hat now denotes the corresponding operators. From the mathematical point of view,
the inverse operator is defined via the integral representation as
\begin{eqnarray}
\label{GF-4}
\frac{i}{\big[i\hat\partial^- + g \hat A^-(x) \big]} =\lim_{\epsilon\to 0}
\int\limits_{0}^{\infty}d\nu  e^{i\nu\big[i\hat\partial^- + g \hat A^-(x) + i\epsilon\big]}.
\end{eqnarray}
Hence, we can write the Green function as
\begin{eqnarray}
\label{GF-5}
{\mathrm S}^{[\gamma^+]}(x,y)&=&i \int\limits_{0}^{\infty}d\nu  e^{i\nu\big[i\hat\partial^- + g \hat A^-(x) + i\epsilon\big]} \delta^{(4)}(x-y)
\nonumber\\
&\equiv& i \int\limits_{0}^{\infty}d\nu {\cal U}(\nu).
\end{eqnarray}
Here and in what follows the limit symbol has been omitted.
In the momentum representation,  ${\cal U}(\nu)$ takes the form of
\begin{eqnarray}
\label{GF-6}
{\cal U}(\nu)=\int(d^4p) e^{-ip(x-y)+i\nu \hat p + i {\cal K}(x,\nu)-\epsilon \nu}
\end{eqnarray}
where the integration measure $(d^4p)$ includes all needed normalization constants and  we use
\begin{eqnarray}
e^{-\nu \hat\partial^-} e^{-ip(x-y)}= e^{-ip(x-y)}  e^{i \nu \hat p^-}
\end{eqnarray}
which defines how the operator acts.
In Eqn.~(\ref{GF-6}), the function ${\cal K}(x,\nu)$ is an unknown function
which we have to derive.

Since the function ${\cal U}(\nu)$ obeys (we can check that by straightforward calculations)
\begin{eqnarray}
\label{GF-7}
-i \frac{\partial {\cal U}(\nu)}{\partial\nu} = \big[i\hat\partial^- + g \hat A^-(x) + i\epsilon\big] {\cal U}(\nu),
\end{eqnarray}
the function ${\cal K}(x,\nu)$ has to satisfy the following differential equation
\begin{eqnarray}
\label{GF-8}
\frac{\partial {\cal K}(x,\nu)}{\partial\nu} = - \partial^- {\cal K}(x,\nu) + g A^-(x)
\end{eqnarray}
provided ${\cal K}(x,\nu=0)=0$.
A solution of Eqn.~(\ref{GF-8}) can be easily found (see, \cite{BogoShir}), it reads
\begin{eqnarray}
\label{GF-9}
{\cal K}(x,\nu)&&= g \int\limits_{0}^{\nu} ds \int(d^4k) e^{-ik(x-s\breve{n}^+)} A^-(k)
\nonumber\\
&&=g\int\limits_{0}^{\nu} ds A^-(x-s \breve{n}^+).
\end{eqnarray}
Using Eqn.~(\ref{GF-9}), the corresponding Green function takes the form of
\begin{eqnarray}
\label{GF-10}
{\mathrm S}^{[\gamma^+]}(x,y) =&&
i \int\limits_{0}^{\infty}d\nu e^{-\epsilon \nu}
\delta^{(4)}(x-y -\nu \breve{n}^+) \times
\nonumber\\
&&{\rm exp}\Big\{-ig\int\limits_{x}^{y} dz^+ A^-(z^+) \Big\}
\end{eqnarray}
where the standard integral representation for $\delta$-function has been used,
\begin{eqnarray}
\delta^{(4)}(x-y -\nu \breve{n}^+)=\int(d^4p) e^{-ip(x-y)+i\nu \breve{n}^+ p},
\end{eqnarray}
and we trade $x-\nu \breve{n}^+$ (see, the upper integral limit in integration over $dz^+$) for $y$
thanks for the argument of $\delta$-function.

The final stage is to write the integration of $\delta$-function as
\begin{eqnarray}
&&i\int\limits_{0}^{\infty} d\nu e^{-\epsilon\nu} \delta^{(4)}(x-y -\nu \breve{n}^+)=
\nonumber\\
&&i\int\limits_{0}^{\infty} d\nu e^{-\nu \hat \partial^- -\epsilon\nu} \delta^{(4)}(x-y)=
\nonumber\\
&&=-\frac{1}{[i\hat\partial^- + i\epsilon]} \delta^{(4)}(x-y)\equiv S^{c\,[\gamma^+]}(x-y).
\end{eqnarray}
Thus, we derive that
\begin{eqnarray}
\label{f-GF-1}
\hspace{-0.5cm}{\mathrm S}^{[\gamma^+]}(x,y)=S^{c\,[\gamma^+]}(x-y)\,[x \,;\, y]_{A^-}
\end{eqnarray}
where $S^c(x-y)$ is defined through $\langle 0| T \psi(x) \bar\psi(y)|0\rangle$ and we use
the obvious property $ [x \,;\, y]_{A}= [y \,;\, x]^{-1}_{A}$.

The extension to the non-Abelian gauge group is straightforward.

From Eqn.~(\ref{f-GF-1}), we can conclude that the fermion field operator in the external field reads
\begin{eqnarray}
\label{f-GF-2}
\hspace{-0.5cm}\Psi(x^+ | A)=\psi(x^+) {\rm exp}\Big\{ig\int\limits_{\mathds{C}}^{x^+} dz^+ A^-(z^+) \Big\},
\end{eqnarray}
where $\mathds{C}$ is, in principle, an arbitrary point which however we choose to be equal to $-\infty^+$.

We stress that the fermion in the external field differs from the gauge-invariant fermion field which appears
in the string operator, see Eqn.~(\ref{Wl-GT-2}).
Indeed, as well-known (see, for example, \cite{Balitsky:1987bk, Radyushkin:1983mj}) in order to get the
gauge-invariant string operator it is necessary to include the gauge boson (gluon) radiations from the fermions
after the interaction of them with photons (or other gauge bosons) as shown in Fig.~\ref{Fig-DY}, the left panel.
Otherwise, we deal with the fermions in the external fields which are {\it not} gauge-invariant
(see, Fig.~\ref{Fig-DY}, the right panel).

To illustrate the last statement, let us consider the simplest case of Compton-like amplitude (see also \cite{Radyushkin:1983mj}).
We have
\begin{eqnarray}
\label{CA-1}
T^{\mu\nu}=\int (d^4x) e^{-iq\cdot x} \langle p | T J^{\mu}(x) J^{\nu}(0)|p \rangle.
\end{eqnarray}
On the handbag diagram level, we have
\begin{eqnarray}
\label{CA-2}
&&T^{\mu\nu}=\int (d^4x) e^{-iq\cdot x}\times
\\
&&\langle p |
: \bar\psi(x)\gamma^{\mu}
\contraction{}{\psi}{(x)}{\bar\psi} \psi(x) \bar\psi(0)
\gamma^{\nu}\psi(0) : |p \rangle.
\nonumber
\end{eqnarray}
In order to include all gauge boson radiations from the fermion propagator given
by the fermion contraction, we merely make a substitution
(modulo the conventional normalizations which are now irrelevant)
\begin{eqnarray}
\label{SubProp-1}
\contraction{}{\psi}{(x)}{\bar\psi} \psi(x) \bar\psi(0) = S^c(x)\Longrightarrow
\mathrm{S}(x,0)
\end{eqnarray}
with $\mathrm{S}(x,0)$ being {\it gauge-noninvariant} Green function, see Eqn.~(\ref{f-GF-1}).
Using the relation which is similar to Eqn.~(\ref{f-GF-1}), we can obtain that
\begin{eqnarray}
\label{CA-3}
&&T^{\mu\nu}=\int (d^4x) e^{-iq\cdot x}\times
\\
&&\langle p |
: \bar\psi(x)\gamma^{\mu}
S^{c}(x)\,[x \,;\, 0]_{A}
\gamma^{\nu}\psi(0) : |p \rangle.
\nonumber
\end{eqnarray}
After the factorization procedure, the matrix combination
$\gamma^{\mu}\,S^{c}\,\gamma^{\nu}$ refers to the so-called hard part,
while the non-perturbative hadron matrix element involves the {\it gauge-invariant}
string operator defined as
\begin{eqnarray}
\langle p |
: \bar\psi(x)\,[x \,;\, 0]_{A} \psi(0) : |p \rangle.
\end{eqnarray}

\section{Gauge and residual gauge symmetries}
\label{G-Rsymmetry:App:B}

In this Appendix, we remind some subtleties related to the residual gauge transformations
in different gauge theories.

\subsection*{Classical $U(1)$-gauge theory (Abelian theory)}

The $U(1)$-gauge theory where the gauge transformation
\begin{eqnarray}
\label{U1}
A_\mu^\Lambda(x)=A_\mu(x) + \partial_\mu \Lambda(x),
\end{eqnarray}
defines an orbit on the $U(1)$-group.
In the Abelian case, the strength  tensor $F_{\mu\nu}$ is gauge-invariant and, therefore,
only the longitudinal (unphysical) components of field, $A^L_\mu$, can be gauge-transformed.
Indeed, in the classical gauge theory for both $k^2=0$ and $k^2\neq 0$,
the solution of the Maxwell equation in vacuum, $\partial_\mu F_{\mu\nu}=0$, reads (modulo the complex conjugated terms),
(see, e.g., \cite{RubakovBook})
\begin{eqnarray}
\label{SolMeq}
&&A_{\mu}(x)=A^L_{\mu}(x) + A^\perp_{\mu}(x)=
\\
&&\int(d^4k)e^{ikx} k_\mu a_L(k) +
\int(d^4k)e^{ikx} \delta(k^2) e^{\perp\,(\alpha)}_\mu a_\perp^{(\alpha)}(k),
\nonumber
\end{eqnarray}
where $(d^4 k)$ stands for the corresponding integration measure with an appropriate normalization,
$\alpha=(1,2)$ and $k\cdot e^{\perp\,(\alpha)}=0$. With this expression, we can easily derive the gauge transformations in
$p$-space ($=$ the momentum representation)
\begin{eqnarray}
\label{GTr-pspace-1}
k_\mu a_{L}^{\Lambda}(k) + e^{\perp}_\mu a_{\perp}^{\Lambda}(k) =
k_\mu a_{L}(k) + e^{\perp}_\mu a_{\perp}(k) + k_\mu\tilde\Lambda(k),
\end{eqnarray}
where the imaginary factor $i$ is absorbed in the definition of $\tilde\Lambda$.
In what follows summation over $\alpha$ and dimensionful normalizations are not shown explicitly
unless it leads to misunderstanding.

Since $k\cdot e^{\perp\,(\alpha)}=0$, we conclude that
\begin{eqnarray}
\label{GTr-pspace-2}
a_{L}^{\Lambda}(k)= a_{L}(k) + \tilde\Lambda(k), \quad a_{\perp}^{\Lambda}(k) = a_{\perp}(k),
\end{eqnarray}
or, equivalently,
\begin{eqnarray}
\label{GTr-pspace-3}
&&A^{L,\,\Lambda}_\mu(k)= A^{L}_\mu(k) + k_\mu \tilde\Lambda(k),
\nonumber\\
&&A^{\perp,\,\Lambda}_\mu(k) = A^{\perp}_\mu(k).
\end{eqnarray}
Moreover, it is easy to demonstrate that
\begin{eqnarray}
\label{AL}
A^{L}_\mu(x) = -i \partial_\mu\alpha(x),
\end{eqnarray}
where $\alpha(x)$ is a scalar function which is related to $a_{L}(k)$ via the Fourier transformation,
$\alpha(x)\stackrel{\text{F}}{=}a_{L}(k)$, and
$a_{L}(k)=\xi(k)/k^2$ with $\xi(k)\stackrel{\text{def}}{=}k\cdot A^{L}(k)\neq 0$
for $k^2\neq 0$. Notice that if $k^2=0$, the Maxwell equation takes the simplest form, $k\cdot A(k)=0$,
in the $p$-space and, therefore, $k\cdot A^{L}(k)= k^2 a_{L}(k)=0$ or, in other words,
$a_{L}(k)=\xi(k)/k^2\sim 0/0$.

As well-known, to fix the certain  representative on the group orbit we have to impose a
gauge condition $F(A^\Lambda)=0$ on the gauge-transformed fields in order to find a solution with respect to
the gauge parameter $\Lambda$. Here, we do not discuss the appearance of Gribov's ambiguity.

{\bf The Lorentz gauge.} As the first example, we consider the Lorentz (covariant) condition which states
\begin{eqnarray}
\label{L-con}
\partial_\mu A^\Lambda_\mu(x)= \partial_\mu A_\mu + \partial^2 \Lambda(x)
=0.
\end{eqnarray}
In $p$-space, the condition (\ref{L-con}) takes the following form:
\begin{eqnarray}
\label{L-con-2}
k_\mu A^L_\mu(k) + k^2 \tilde\Lambda(k)=0
\end{eqnarray}
which gives us the relation $a_L(k)=-\tilde\Lambda(k)$ for the case of $k^2\neq 0$.
Notice that if $k^2=0$, then the functions $a_L(k)$ and $\tilde\Lambda(k)$ in the
combination $a_L(k)+\tilde\Lambda(k)$ are free functions and they are independent of each other.

However, the gauge condition (\ref{L-con}) (or (\ref{L-con-2})) can not fix the orbit
representative uniquely. Indeed, there is still the so-called residual gauge freedom
defined by $F(A^\Lambda)=F(A)=0$. For the Lorentz condition,
two simultaneous conditions:
\begin{eqnarray}
\label{ResG-1}
\partial_\mu A^\Lambda_\mu(x)=0\quad\text{and}\quad \partial_\mu A_\mu(x)=0
\end{eqnarray}
lead to
\begin{eqnarray}
\label{ResG-2}
\partial^2 \Lambda_0(x)=0
\end{eqnarray}
where the gauge function (parameter) $\Lambda_0$ defines the residual gauge freedom.
That is, the residual gauge transformation with the function $\Lambda_0$ keeps the gauge
condition, $F(A)=0$, gauge-invariant.  Hence, the gauge freedom fixing
means that one fixes all gauge freedom including the residual gauge.
In other words, if there is no residual gauge transformation, the
given gauge condition fixes the gauge freedom completely and we deal with
one representative on a gauge orbit.

Let us consider the second gauge condition in Eqn.~ (\ref{ResG-1}). In $p$-space, it leads to
the following possibilities ($k\cdot a_\perp=0$ by definiton)
\begin{eqnarray}
\label{L-con-p-1}
k^2 a_L(k)=0\,\, \Longrightarrow
\begin{cases}
k^2=0, &a_L(\vec{k})-\,\text{arbitrary }\,\\
k^2\neq 0, &a_L(k)=0.\\
\end{cases}
\end{eqnarray}
Hence, we can see that the gauge condition (\ref{ResG-1}) cannot eliminate the
unphysical field $A^L_\mu$ for the case of $k^2=0$. Working with the equation
(\ref{ResG-2}), in the same manner, we conclude that the gauge function
$\tilde\Lambda_0(\vec{k})$ is not fixed and generates the residual gauge transformation provided
$k^2=0$.

It is instructive to consider the condition (\ref{ResG-2}) in the coordinate representation ($x$-space).
Solutions (\ref{ResG-2}) can be easily found and represented, for instance, the following form:
\begin{eqnarray}
\label{Sol-ReG-x}
\Lambda_0(x)=\begin{cases}
\text{const} &\,\\
1/x^2, &\text{for}\,\,\, x^2\neq 0\,\\
C_0 e^{i(x_0-\vec{x}\vec{N})}&\text{with},\,\,\, |\vec{N}|=1.\\
\end{cases}
\end{eqnarray}
Notice that the scalar function $\alpha(x)$ in Eqn.~(\ref{AL}) which obeys the second condition in Eqn.~ (\ref{ResG-1}),
{\it i.e.} $\partial^2\alpha(x)=0$,  has formally the same form as (\ref{Sol-ReG-x}).

For $k^2\neq0$, the scalar gauge function $\Lambda$ gives also
the longitudinal (unphysical) field $A^L_\mu$, see (\ref{L-con-2}). Therefore,
the first two solutions of (\ref{Sol-ReG-x}) are irrelevant for our study.
In order to get matched with the corresponding condition (\ref{L-con-p-1}) in the momentum representation,
we have to put $C_0$ be equal to zero, $C_0=0$. However, for the case of $k^2=0$,
as above-mentioned, the functions $\alpha(x)$ and $\Lambda_0(x)$ are independent and arbitrary
due to the different free constant pre-factors in the plane wave solution.

We can also consider the Lorentz gauge condition (\ref{L-con}) as an inhomogeneous
differential equation with respect to $\Lambda(x)$, {i.e.}
\begin{eqnarray}
\label{L-con-de}
\partial^2\Lambda(x)=\eta(x)
\end{eqnarray}
where $\eta(x)\stackrel{\text{def}}{=}-\partial_\mu A_\mu(x)$. Solving (\ref{L-con-de}), we obtain that
\begin{eqnarray}
\label{Sol-L-con-de}
\Lambda(x)=\Lambda_0(x)+\int d^4y \,G(x-y) \eta(y),
\end{eqnarray}
where the Green function $G(x)$ is defined as
\begin{eqnarray}
G(x)=\frac{1}{[\partial^2]_{\text{reg}}}\delta^{(4)}(x)
\end{eqnarray}
with the suitable regularization of operator stemmed from
the corresponding boundary conditions, see \cite{BogoShir}.

{\bf The Coulomb gauge.} Using the condition $A_0^\Lambda(x)=0$ to amplify the Lorentz condition (\ref{L-con}),
we can get the Coulomb gauge condition which reads
\begin{eqnarray}
\label{C-con}
\vec{\partial}\vec{A}^\Lambda(x)= \vec{\partial}\vec{A}(x) + \Delta\Lambda(x)=0.
\end{eqnarray}
In $p$-space, the condition (\ref{C-con}) is transformed to (recall that $\vec{\partial}\vec{A}^\perp=0$
by construction)
\begin{eqnarray}
\label{C-con-p}
\vec{k}^2 a_L(k) + \vec{k}^2 \tilde\Lambda(k) = 0.
\end{eqnarray}
Again, let us study the corresponding residual gauge freedom:
\begin{eqnarray}
\label{RegC-con-1}
\vec{\partial}\vec{A}(x)=0\quad\text{and}\quad  \Delta\Lambda(x)=0.
\end{eqnarray}
For the sake of simplicity, we dwell on the case of $k^2=0$ which leads to $\vec{k}^2\neq 0$. With this,
instead of (\ref{C-con-p}), it is enough to stop on the equation
\begin{eqnarray}
\label{C-con-L}
\vec{k}^2 \tilde\Lambda(\vec{k})=0.
\end{eqnarray}
Hence,  the only solution of (\ref{C-con-L}) is $\tilde\Lambda=0$ which means that there is no any
residual freedom at all.

Therefore, in the Coulomb gauge there are no the longitudinal field components and we deal with the
physical gauge field $A^\perp_\mu$ only.

{\bf The Hamilton and axial gauges.} In the similar manner, we can study the residual gauge symmetries in
the Hamilton ($A_0^\Lambda=0$) and axial ($A^{+,\,\Lambda}=0$) gauges. The residual gauge transformations
are given by the corresponding free (unfixed) gauge function $\tilde\Lambda(k)$ provided
$k_0=0$ or $k^+=0$.

\subsection*{Classical $SU(3)$-gauge theory (Non-abelian theory)}

The next subject of our discussion is a non-Abelian gauge theory with
$SU(3)$ gauge group. In this case  case, the gauge transformation is given by
\begin{eqnarray}
\label{SU3}
A^\omega_\mu(x)=\omega(x) A_\mu(x) \omega^{-1}(x) +
\frac{i}{g}\omega(x)\partial_\mu \omega^{-1}(x)
\end{eqnarray}
which gives in the infinitesimal form
\begin{eqnarray}
\label{SU3inf}
A^{a,\,\omega}_\mu(x)= A^a_\mu(x) +
f^{abc} A^b_\mu(x) \theta^c(x) +
\frac{1}{g} \partial_\mu \theta^a(x),
\end{eqnarray}
where $\omega(x)=\exp\big( i \theta^a(x) t^a\big)$.
The decomposition of field components in the longitudinal and transverse
components is similar to the Abelian case, see above.
In contrast to the $U(1)$ gauge group, the strength tensor $G_{\mu\nu}$
is gauge-covariant. It means that all field components may change under
gauge transformations.

{\bf The Lorentz gauge.} We again begin with the Lorentz gauge condition:
\begin{eqnarray}
\label{SU3inf-L-con}
&&\partial_\mu A^{a,\,\omega}_\mu(x)=
\\
&&\partial_\mu A^a_\mu(x) +
f^{abc} \partial_\mu \big( A^b_\mu(x) \theta^c(x)\big) +
\frac{1}{g} \partial^2 \theta^a(x)=0.
\nonumber
\end{eqnarray}
As above-mentioned, the gauge condition is invariant under
the residual gauge transformation:
\begin{eqnarray}
\label{ReL-con-SU-1}
\partial_\mu A^{a,\,\omega}_\mu(x)=\partial_\mu A^a_\mu(x)=0
\end{eqnarray}
or, equivalently,
\begin{eqnarray}
\label{ReL-con-SU-2}
{\cal D}_\mu^{ac}\partial_\mu \theta^c(x)=0,
\end{eqnarray}
where ${\cal D}_\mu^{ac}=\partial_\mu\delta^{ac} + g f^{abc} A^b_\mu(x)$.

In $p$-space, the condition (\ref{ReL-con-SU-2}) takes the form of
\begin{eqnarray}
\label{ReL-con-SU-p-1}
-k^2 \theta^a(k) + i g f^{abc} k_\mu A^{b,\,L}_\mu(k) \theta^c(k)=0.
\end{eqnarray}
If $k^2=0$ and, therefore, $k_\mu A^{b,\,L}_\mu(k)=0$, then the gauge function $\theta(x)$ cannot be fixed
and generates the residual gauge transformation.

{\bf The Hamilton gauge.} The similar situation occurs in the Hamilton gauge,
$A^\omega_0=0$. The residual transformation is induced by the gauge function which obeys
\begin{eqnarray}
\label{ReH-con}
\partial_0 \theta^a(x_0,\vec{x})=0.
\end{eqnarray}
Hence, the solution of this equation is rather trivial: $\theta$-function is the time-independent function, $\theta_0(\vec{x})$.

In the momentum representation, the condition (\ref{ReH-con}) gives us the equation
\begin{eqnarray}
\label{ReH-con-p-1}
\int (d^4k)e^{ikx} k_0 \theta^a(k_0,\vec{k})=0
\end{eqnarray}
which has a solution as
\begin{eqnarray}
\label{Sol-ReH-con-p}
\theta^a_0(k)=\delta(k_0)\theta^a_0(\vec{k}).
\end{eqnarray}
Therefore, we find in the coordinate representation
\begin{eqnarray}
\int (d^4k)e^{ikx} \delta(k_0)\theta^a_0(\vec{k}) = \theta^a_0(\vec{x})
\end{eqnarray}
which coincides with the results of the preceding paragraph.

{\bf The axial gauge.} Working in the axial gauge, $A^{+,\,\omega}=0$, in the similar manner
we are able to find the gauge function that is responsible for the residual gauge symmetry.
We impose the condition
\begin{eqnarray}
\label{RecAx-con-1}
A^{+,\,\omega}(x)=A^{+}(x)=0
\end{eqnarray}
or, in the equivalent form,
 \begin{eqnarray}
\label{RecAx-con-2}
\partial^{+}\theta^a(x^+,x^-,\vec{\bf x}_\perp)=0\quad \text{with}\quad
\partial^+=\partial_-=\frac{\partial}{\partial x^-}.
\end{eqnarray}
The solution of this trivial differential equation is the $x^-$-independent function
$\theta^a_0(x^+,\vec{\bf x}_\perp)$ which has the following form in $p$-space
(cf. (\ref{ReH-con-p-1}) and (\ref{Sol-ReH-con-p})):
\begin{eqnarray}
\label{ReAx-con-p-1}
\theta^a_0(k^+,k^-,\vec{\bf k}_\perp)=\delta(k^+)\theta^a_0(k^-,\vec{\bf k}_\perp)
\end{eqnarray}
where $\theta^a_0(k^-,\vec{\bf k}_\perp)$ is an arbitrary gauge function related to the residual symmetry.

It is instructive to focus on the finite gauge transformations and corresponding gauge condition, namely
\begin{eqnarray}
\label{ReAx-f-con}
&&A^{+,\,\omega}(x)=
\\
&&\omega(x) A^+(x) \omega^{-1}(x) +
\frac{i}{g}\omega(x)\partial^+ \omega^{-1}(x)=0.
\nonumber
\end{eqnarray}
The solution of this equation can easy be found, it reads
\begin{eqnarray}
\label{Sol-Ax-con-f}
\omega_0(x)=\mathbb{P}{\rm exp}\Big\{ ig \int\limits^{x^-}_{\mathds{C}}
dz^- A^+(x^+,z^-,\vec{\bf x}_\perp) \Big\}
\end{eqnarray}
where, generally speaking, $\mathds{C}$ is an arbitrary constant.
We stress that the solution $\omega_0(x)$ is valid for $\forall x \in \mathds{R}^4$.
At the same time, this function can be multiplied by an arbitrary $x^-$-independent
gauge function to produce another solution of equation (\ref{ReAx-f-con}), {\it i.e.}
\begin{eqnarray}
\label{Omega-gen}
W(x^+,x^-,\vec{\bf x}_\perp) = \bar\omega(x^+,\vec{\bf x}_\perp) \omega_0(x^+,x^-,\vec{\bf x}_\perp),
\end{eqnarray}
where $\bar\omega(x^+,\vec{\bf x}_\perp)=\exp\big( i\theta^a(x^+,\vec{\bf x}_\perp)t^a \big)$.
Indeed, one can demonstrate that $A^{+,\,W}(x)=0$.

To study the residual symmetry, we have to demand that $A^+(x)=0$ for any $x$.
Therefore, from (\ref{Omega-gen}), we obtain that the function
\begin{eqnarray}
\label{Omega-gen-res}
W(x^+,x^-,\vec{\bf x}_\perp)\Big|_{A^+=0} = \bar\omega(x^+,\vec{\bf x}_\perp)
\end{eqnarray}
generates the residual transformation we are interested in.

Let us now return to the gauge function presented by (\ref{ReAx-con-p-1}).
The case of $k^+=0$ (which provides us the residual symmetry) leads to the
so-called spurious singularity in the gluon propagator in the axial gauge, see the next subsection.
If we adopt a procedure to regularize this singularity with the help of some well-defined procedure,
$[ k^+]_{reg}\neq 0$, then the existence condition for the residual symmetry, see (\ref{RecAx-con-2}),
has to be given by (in the momentum representation)
\begin{eqnarray}
\label{ReAx-con-reg}
\int (d^4k)e^{ikx} [k^+]_{reg} \delta(k^+)\theta^a_0(k^-,\vec{\bf k}_\perp)=0.
\end{eqnarray}
Hence, the only possibility to satisfy this equation is to demand that $\theta^a_0(k^-,\vec{\bf k}_\perp)=0$ which
means that we fix the remaining residual symmetry.
Thus, we conclude that the spurious singularity is fixed if and only if we do not have the residual gauge symmetry.
On the other hand, we may say that the residual gauge fixing is enough for the elimination of spurious singularity.

\subsection*{Spurious singularity of gluon propagator}

Let us return to the issue of the spurious singularity which appears in the gluon propagator in
the axial gauge $A^+=0$.

The generating functional for gluons (gluonodynamics)
in the most general gauge $F(A^\theta)=0$
\begin{eqnarray}
\label{Z}
&&\mathbb{Z}=N \int {\cal D} A_\mu e^{iS[A]} =
\\
&&\tilde N \int {\cal D} A_\mu \, \Delta_c[A]\, \delta\big(F(A)\big) \, e^{iS[A]},
\nonumber
\end{eqnarray}
where $\tilde N$ involves the infinite gauge group volume, $\int d\theta$, and we use
\begin{eqnarray}
\mathds{1}= \int d\theta \Delta_c[A] \delta\big( F(A^\theta)\big),
\quad \Delta_c[A^\theta] =  \Delta_c[A].
\end{eqnarray}
Instead of solving the gauge condition $F(A^\theta)=0$ with respect to the group function
$\theta$ within the generalized Hamilton formalism, we separate out the infinite group volume, $\int d\theta$,
in the generating functional (the Faddeev-Popov approach).

The next trick is related to the exponentiation of $\delta\big(F(A)\big)$. We introduce the generalized gauge condition as
$F(A)=C$ with $\delta C/\delta A_\mu=0$.  The generalizing functional $\mathbb{Z}$ must be independent on $C$.
Therefore, to get the $C$-independent functional we have to integrate out over this parameter $C$.
Using the integration measure defined as
\begin{eqnarray}
dC \exp\big( - \frac{i}{2\xi} \int d^4 x \,  C^2(x) \big),
\end{eqnarray}
we have
\begin{eqnarray}
\label{Z-2}
&&\hspace{-0.5cm}\mathbb{Z}=\tilde N \int dC e^{\big( - \frac{i}{2\xi} \int d^4 x C^2(x) \big)}
\int {\cal D} A_\mu \, \Delta_c[A]\, \delta\big(F(A)-C\big) \, e^{iS[A]}
\nonumber\\
&&=\tilde N
\int {\cal D} A_\mu \, \Delta_c[A] \, e^{iS[A]- \frac{i}{2\xi} \int d^4 x F^2(A)}.
\end{eqnarray}
In (\ref{Z-2}), the effective action  with the gauge-fixing term,
\begin{eqnarray}
S_{\text{fix}}=- \frac{1}{2\xi} \int d^4 x F^2(A),
\end{eqnarray}
is now not gauge-invariant anymore. As a result of this trick, we don't need to solve
the gauge condition with respect  to the gauge function.

Let the gauge condition $F(A)=0$ be $A^+=0$ with $n^2=0$. In this case, the determinant $\Delta_c[A]$ is independent on $A$
and, therefore, we are able to include this determinant in the normalization of functional.
Thus, the effective Lagrangian reads
\begin{eqnarray}
\label{Leff}
{\cal L}_{\text{eff}}= -\frac{1}{4} G_{\mu\nu}G_{\mu\nu} - \frac{1}{2\xi} \big( n\cdot A \big)^2.
\end{eqnarray}
This Lagrangian yields the effective action which can be written as
\begin{eqnarray}
\label{Seff}
S_{\text{eff}}= \frac{1}{2} \int d^4x\, A_\mu(x)\, K_{\mu\nu}(x)\, A_\nu(x),
\end{eqnarray}
where
\begin{eqnarray}
\label{Kop}
K_{\mu\nu}(x)=g_{\mu\nu}\partial^2 - \partial_\mu \partial_\nu - \frac{1}{\xi} n_\mu n_\nu.
\end{eqnarray}
In $p$-space, the operator $K_{\mu\nu}$ has an inverse operator which, in the limit of $\xi\to 0$, is given by
\begin{eqnarray}
\label{InverseK}
&&K^{-1}_{\mu\nu}(k)=\frac{d_{\mu\nu}(k,n)}{k^2+i0},\quad
\nonumber\\
&&d_{\mu\nu}(k,n)=g_{\mu\nu}-\frac{k_\mu n_\nu + k_\nu n_\mu}{k^+}.
\end{eqnarray}
As we have demonstrated in the preceding subsection, when we fix/regularize
the spurious singularity $[k^+]_{reg}$ it means that we fix the residual gauge
symmetry defined by the gauge function $\theta^a(k^-,\vec{\bf k}_\perp)$ and {\it vise versa}.

We also remind that it is not possible to fix the residual gauge simply by means of adding of
\begin{eqnarray}
\label{Amin-Leff}
\frac{1}{2\xi_2}\big( n^*\cdot A \big)^2
\end{eqnarray}
in Eqn.~(\ref{Leff}). In this case, the inverse kinematical operator (see, Eqn.~(\ref{InverseK})) does not
exist due to the fact that the free (without the coefficients) tensors $n_{\mu}n_{\nu}$ and
$n^*_{\mu}n^*_{\nu}$ present in the corresponding equation to determine the coefficients.
Indeed, introducing the Lorentz parametrization (where the coefficients have to be determined)
\begin{eqnarray}
\label{Lor-par-d}
&&d_{\nu\rho}(k,n,n^*)=g_{\nu\rho} + a_1 k_{\nu}k_{\rho}+
b_2 k_{\nu}n_{\rho}+
b_3 n_{\nu}k_{\rho}+
\nonumber\\
&&b_4 k_{\nu}n^*_{\rho}+
b_5 n^*_{\nu}k_{\rho}+
c_6 n_{\nu}n_{\rho}+
c_7 n^*_{\nu}n^*_{\rho},
\end{eqnarray}
where
\begin{eqnarray}
\text{dim}_M[a_1]=-2,\, \text{dim}_M[b_i]=-1,\, \text{dim}_M[c_j]=0,
\end{eqnarray}
the contraction equation on the coefficients (or, in other words, the Green function equation)
\begin{eqnarray}
\label{Cont-Eqn}
K_{\mu\nu}d_{\nu\rho}=g_{\mu\rho}
\end{eqnarray}
involves the tensors $n_{\mu}n_{\nu}$ and $n^*_{\mu}n^*_{\nu}$ which stay without
coefficients. It means that the inverse operator cannot be derived.


\begin{thebibliography}{99}

\bibitem{Angeles-Martinez:2015sea}
  R.~Angeles-Martinez {\it et al.},
  Acta Phys.\ Polon.\ B {\bf 46} (2015) no.12,  2501
  [arXiv:1507.05267 [hep-ph]].
\bibitem{Boer:2011fh}
  D.~Boer {\it et al.},
  arXiv:1108.1713 [nucl-th].

\bibitem{Boer:2003cm}
  D.~Boer, P.~J.~Mulders and F.~Pijlman,
  Nucl.\ Phys.\ B {\bf 667}, 201 (2003)

\bibitem{Kang:2011hk}
  Z.~B.~Kang, J.~W.~Qiu, W.~Vogelsang and F.~Yuan,
  Phys.\ Rev.\ D {\bf 83}, 094001 (2011)

\bibitem{Boer:2011fx}
  D.~Boer,
  Phys.\ Lett.\ B {\bf 702}, 242 (2011)

\bibitem{P-R}
  B.~Pire and J.~P.~Ralston,
  Phys.\ Rev.\  D {\bf 28}, 260 (1983).

\bibitem{Carlitz:1992fv}
  R.~D.~Carlitz and R.~S.~Willey,
  Phys.\ Rev.\  D {\bf 45}, 2323 (1992).

\bibitem{Brandenburg:1995pk}
  A.~Brandenburg, D.~Mueller and O.~V.~Teryaev,
  Phys.\ Rev.\  D {\bf 53}, 6180 (1996)

\bibitem{Bakulev:2007ej}
  A.~P.~Bakulev, N.~G.~Stefanis and O.~V.~Teryaev,
  Phys.\ Rev.\  D {\bf 76}, 074032 (2007)

\bibitem{Radyushkin:2009zg}
  A.~V.~Radyushkin,
  Phys.\ Rev.\  D {\bf 80}, 094009 (2009)

\bibitem{Polyakov:2009je}
  M.~V.~Polyakov,
  JETP Lett.\  {\bf 90}, 228 (2009)

\bibitem{Mikhailov:2009sa}
  S.~V.~Mikhailov and N.~G.~Stefanis,
  Mod.\ Phys.\ Lett.\  A {\bf 24}, 2858 (2009)

\bibitem{Hammon:1996pw}
  N.~Hammon, O.~Teryaev and A.~Schafer,
  Phys.\ Lett.\ B {\bf 390}, 409 (1997)

\bibitem{Boer:1997bw}
  D.~Boer, P.~J.~Mulders and O.~V.~Teryaev,
  Phys.\ Rev.\ D {\bf 57}, 3057 (1998)

\bibitem{AT-GP}
 I.~V.~Anikin and O.~V.~Teryaev,
  Eur.\ Phys.\ J.\ C {\bf 75} (2015) no.5,  184;
  Phys.\ Lett.\ B {\bf 751} (2015) 495;
  Phys.\ Lett.\ B {\bf 690} (2010) 519.

\bibitem{Belitsky:2002sm}
  A.~V.~Belitsky, X.~Ji and F.~Yuan,
  Nucl.\ Phys.\ B {\bf 656} (2003) 165

\bibitem{Chirilli:2015fza}
  G.~A.~Chirilli, Y.~V.~Kovchegov and D.~E.~Wertepny,
  JHEP {\bf 1512}, 138 (2015)

\bibitem{Soper:1979fq}
  D.~E.~Soper,
  Phys.\ Rev.\ Lett.\  {\bf 43} (1979) 1847.

  \bibitem{Collins:1981uw}
  J.~C.~Collins and D.~E.~Soper,
  Nucl.\ Phys.\ B {\bf 194} (1982) 445.

\bibitem{Leibbrandt:1987qv}
  G.~Leibbrandt,
  Rev.\ Mod.\ Phys.\  {\bf 59} (1987) 1067.

  \bibitem{Slavnov:1987yh}
  A.~A.~Slavnov and S.~A.~Frolov,
  Theor.\ Math.\ Phys.\  {\bf 73} (1987) 1158
   [Teor.\ Mat.\ Fiz.\  {\bf 73} (1987) 199].

  \bibitem{Bassetto:1991ue}
  A.~Bassetto, G.~Nardelli and R.~Soldati,
  {\it ``Yang-Mills theories in algebraic noncovariant gauges: Canonical quantization and renormalization,''}
  Singapore, Singapore: World Scientific (1991) 227 p

  \bibitem{Bassetto:1984dq}
  A.~Bassetto, M.~Dalbosco, I.~Lazzizzera and R.~Soldati,
  Phys.\ Rev.\ D {\bf 31} (1985) 2012.

  \bibitem{Collins:1989gx}
  J.~C.~Collins, D.~E.~Soper and G.~F.~Sterman,
  Adv.\ Ser.\ Direct.\ High Energy Phys.\  {\bf 5} (1989) 1,

  \bibitem{Ivanov:1990vy}
  S.~V.~Ivanov,
  Fiz.\ Elem.\ Chast.\ Atom.\ Yadra {\bf 21} (1990) 75.

\bibitem{Stefanis:1983ke}
  N.~G.~Stefanis,
  Nuovo Cim.\ A {\bf 83} (1984) 205.

\bibitem{Barone}
  V.~Barone, A.~Drago and P.~G.~Ratcliffe,
  Phys.\ Rept.\  {\bf 359}, 1 (2002)

\bibitem{Efremov:1984ip}
  A.~V.~Efremov and O.~V.~Teryaev,
  Phys.\ Lett.\  B {\bf 150}, 383 (1985).

\bibitem{Anikin:2009bf}
  I.~V.~Anikin, D.~Y.~Ivanov, B.~Pire, L.~Szymanowski and S.~Wallon,
  Nucl.\ Phys.\ B {\bf 828}, 1 (2010)

\bibitem{An}
  I.~V.~Anikin and O.~V.~Teryaev,
  Phys.\ Part.\ Nucl.\ Lett.\  {\bf 6}, 3 (2009)

\bibitem{Braun}
  V.~M.~Braun, D.~Y.~Ivanov, A.~Schafer and L.~Szymanowski,
  Nucl.\ Phys.\  B {\bf 638}, 111 (2002)

\bibitem{BogoShir}
  N.~N.~Bogolyubov and D.~V.~Shirkov,
  ``Introduction To The Theory Of Quantized Fields,''
  Intersci.\ Monogr.\ Phys.\ Astron.\  {\bf 3}, 1 (1959).

\bibitem{Efremov:1978xm}
  A.~V.~Efremov and A.~V.~Radyushkin,
  Theor.\ Math.\ Phys.\  {\bf 44}, 774 (1981)
  [Teor.\ Mat.\ Fiz.\  {\bf 44}, 327 (1980)].

\bibitem{ContourG}
  S.~V.~Ivanov, G.~P.~Korchemsky and A.~V.~Radyushkin,
  Yad.\ Fiz.\  {\bf 44}, 230 (1986)
  [Sov.\ J.\ Nucl.\ Phys.\  {\bf 44}, 145 (1986)];
  S.~V.~Ivanov and G.~P.~Korchemsky,
  Phys.\ Lett.\  B {\bf 154}, 197 (1985);

\bibitem{Vladimirov}
V.~S.~Vladimirov,
``Generalized Functions in Mathematical Physics'',
Mir Publishers (1979) 390 p.

\bibitem{Gross:1971wn}
  D.~J.~Gross and S.~B.~Treiman,
  Phys.\ Rev.\ D {\bf 4}, 1059 (1971).

\bibitem{Balitsky:1987bk}
  I.~I.~Balitsky and V.~M.~Braun,
  Nucl.\ Phys.\ B {\bf 311}, 541 (1989).

\bibitem{Radyushkin:1983mj}
  A.~V.~Radyushkin,
  Fiz.\ Elem.\ Chast.\ Atom.\ Yadra {\bf 14}, 58 (1983).


\bibitem{RubakovBook}
  V.~A.~Rubakov,
  ``Classical theory of gauge fields'',
  Princeton, USA: Univ. Pr. (2002) 444 p.
\end{thebibliography}
\end{document}